\documentclass{article}
\usepackage{amssymb}
\usepackage{amsfonts}
\usepackage{amsmath}
\usepackage[english,german]{babel}
\usepackage{latexsym}
\usepackage{color}
\usepackage[pdftex]{graphicx}
\usepackage{subfigure}
\usepackage[onehalfspacing]{setspace}
\usepackage{amscd}
\usepackage{amsthm}
\usepackage{thmtools, thm-restate}
\usepackage{multirow}
\usepackage{booktabs}
\usepackage{rotating}
\usepackage{layout}
\usepackage{natbib}

\declaretheorem[numberwithin=section]{thm}

\newtheorem{theorem}[thm]{Theorem}
\newtheorem{proposition}[thm]{Proposition}
\newtheorem{corollary}[thm]{Corollary}
\newtheorem{lemma}[thm]{Lemma}

\newtheorem{remark}[thm]{Remark}
\newtheorem{example}[thm]{Example}
\newtheorem{Assumption}[thm]{Assumption}

\newenvironment{Proof}{\textsc{Proof.}}{\mbox{ } \hfill $\Box$ \vspace{2mm}}

\def\be{\begin{equation*}}
\def\ee{\end{equation*}}

\newcommand{\ba}{\begin{array}{ll}}
\newcommand{\bal}{\begin{array}{ll}}
\newcommand{\ea}{\end{array}}

\newcommand{\n}{\mathbb{N}}

\newcommand{\re}{\mathbb{R}}

\newcommand{\C}{\mathcal{C}}

\def\T{\Theta}
\def\t{\theta}

\def\i{{\mbox{\rm{1}\hspace{-0.09in}\rm{1}\hspace{0.00in}}}}

\begin{document}
\selectlanguage{english}

\begin{center}\begin{large}\textbf{A Principal--Agent Model of Trading Under Market Impact\footnote{The research leading to these results has received funding from the ERC (grant agreement 249415-RMAC), from the Swiss Finance Institute project {\sl Systemic Risk and Dynamic Contract Theory}, as well as the SFB 649 {\sl Economic Risk}, and it is gratefully acknowledged.} \\ -Crossing networks interacting with dealer markets-}\end{large}

\vspace{.2cm}
Jana Bielagk\footnote{Department of Mathematics, Humboldt-University Berlin,  Unter den Linden 6, 10099 Berlin, Germany. \\ \hspace{1cm} bielagk@math.hu-berlin.de}, 
Ulrich Horst\footnote{Department of Mathematics, Humboldt-University Berlin,  Unter den Linden 6, 10099 Berlin, Germany. \\horst@math.hu-berlin.de} \& 
Santiago Moreno--Bromberg\footnote{ Center for Finance and Insurance, Department of Banking and Finance, University of Zurich, Plattenstr. 14, 8032 Zurich, Switzerland. santiago.moreno@bf.uzh.ch}
\end{center}

%
%




\begin{abstract} We use a principal--agent model to analyze the structure of a book--driven \textit{dealer market} when the dealer faces competition from a crossing network or dark pool.
The agents are privately informed about their \textit{types} (e.g. their portfolios), which is something that the dealer must take into account when engaging his counterparties. Instead of trading with the dealer, the agents may chose to trade in a \textit{crossing network}. We show that the presence of such a network results in more types being serviced by the dealer
and that, under certain conditions and due to reduced adverse selection effects, the book's \textit{spread} shrinks. We allow for the pricing on the dealer market to
determine the structure of the crossing network and show that the same conditions
that lead to a reduction of the spread imply the existence of an equilibrium
book/crossing network pair.

\medskip


\end{abstract}



\section{\large{Introduction}}

Recently, the analysis of optimal trading under market impact has received considerable attention. Starting with the  contribution of~\cite{AlmgrenChriss00}, the existence of optimal trading strategies under illiquidity has been established by many authors, including ~\cite{Forsyth2012}, \cite{GatheralSchied11}, \cite{KratzSchoeneborn13} and~\cite{SchiedSchoenebornTehranchi10}, just to name a few. The literature on trading under illiquidity typically assumes that block trading takes place under some (exogenous) pricing schedule, which describes the liquidity available for trading at different price levels.	
This article studies the impact of a CN on a DM within the scope of principal--agent models under hidden information (adverse selection). This asymmetric--information approach is a significant departure from the settings of the articles mentioned above. Specifically, we consider a one--period model where block trading is modeled via a risk--neutral dealer or market--maker who provides liquidity to a heterogeneous (in terms of idiosyncratic characteristics or ``types'') group of privately--informed investors or traders. Extending the seminal work on asset pricing under asymmetric information in~\cite{BMR}, we assume that each investor has an outside option that provides him with a type--dependent reservation utility that the dealer may not be able to match without making a loss. We allow the dealer to abstain from trading with investors whose outside options would be too costly to match. The fact that the dealer may choose between excluding agents, matching their outside options (which in some cases yields him strictly positive profits) or offering them contracts that result in utilities that strictly dominate their reservation ones, implies that a rich structure (in terms of the partition of the type space) may emerge in equilibrium. For instance, in Example~\ref{RichStructure} we analyze a scenario where the type space is partitioned into two intervals where the agents' outside options are matched, one where they are excluded and three where they earn positive rents. In more mundane terms, within a portfolio--liquidation framework, we may think of traders who need to unwind portfolios whose sizes are private information and who can either trade in a DM or a CN, the latter providing some of them with trading options that the dealer may be unable improve upon without suffering losses. To the best of our knowledge, such adverse--selection models have thus far only been considered by~\cite{BJ:03} and~\cite{Page}. 
The latter analyzes, in quite a general setting where the set of consumer types is a Polish space and the contract space an arbitrary compact metric space,
the problem  of a monopolist who faces both an adverse-selection problem (as in the work at hand) as well as a moral-hazard one relative to contract performance. 
\cite{BJ:03}, on the other hand, only studies the adverse-selection problem in a finite-dimensional setting. This allows him to find a quasi-explicit representation of the optimal contract using Lagrange-multiplier techniques. He identifies conditions for the optimal contract to be separating, to be non--stochastic and to induce full participation. Furthermore, he also discusses the nature of the solution when bunching occurs. He does not, however, analyze the case where the dealer's choices may have an impact on the structure of the reservation--utility function, which in turn would influence his decisions. Our study of such a feedback loop is novel and it is a crucial component in our analysis of the interactions between DMs and CNs, which is typically not unidirectional. To account for the fact that many off--exchange venues settle trades at prices taken from primary venues, we state sufficient conditions for the existence of an equilibrium pricing schedule. By this we mean that there exists a pricing schedule in the DM such that, if trades in the CN are settled at the best bid and ask prices from the DM, then the dealer's optimal pricing schedule is precisely that schedule.

\vspace{0.2cm}
In order to study the impact of a type--dependent outside option, we first analyze the benchmark case where the said option is trivial, i.e. all traders may abstain from engaging the dealer and in turn earn (or lose) nothing. In such a setting the dealer is able to match the traders' outside options by offering ``nothing in exchange for nothing'', which is costless. This analysis follows~\cite{BMR}. Next we look at the general case where the traders' reservation utilities are type dependent and the dealer need not be able to match them without incurring losses. It is well known that asymmetric information results, in equilibrium, in some traders being kept to their reservation utilities. This is due to the adverse--selection costs. Intuitively, these costs increase with the profitability of trading with high--type traders (e.g. investors with large portfolios). This suggests that when mostly high--type traders benefit from the outside option in terms of the latter strictly dominating what the dealer would have offered them in the benchmark case, then more low--type traders will be serviced in equilibrium. As a consequence of the reduced adverse--selection costs, more investors engage in trading, either in the DM or the CN. Our analysis further suggests that the presence of the CN is welfare improving even for investors for whom trading in the CN is not beneficial. We also provide sufficient conditions that guarantee that the competition from the CN results in a narrower spread in the DM. Overall, we propose a benchmark model of optimal block trading of privately--informed traders with an endogenous pricing schedule, analyze the impact of a CN on pricing schedules in DMs and prove an existence result of equilibria of best bid and ask prices in our trading game.

\vspace{0.2cm}
\subsection*{Related literature}	
\cite{HorstNaujokat13} and~\cite{KratzSchoeneborn13} were the first to allow orders to be simultaneously submitted both to a dealer market (DM) and to an off--exchange venue such as a crossing network (CN) or a dark pool (DP). These are alternative trading facilities that allow investors to reduce their market impact by submitting liquidity that is shielded from the public view. The downside is that trade execution is uncertain: trades take place only when the matching liquidity is or becomes available. In such a case, trades are typically settled at prices prevailing in an associated primary venue, which significantly reduces the cost of large trades if settled in a CN or in a DP.	The aforementioned articles on optimal, simultaneous trading in DMs and CNs do not allow for an impact of off--exchange trading on the dynamics of the associated DM. Equilibrium models analyzing the impact of alternative trading venues on DMs and trading behavior have been extensively analyzed in the financial--economics literature; see, e.g.~\cite{Glosten} and~\cite{PS} and the references therein. To simplify the analysis of market impact, this literature typically assumes that the market participants trade only a single unit of the stock. For instance, in their seminal work, \cite{HM} derives conditions for the viability of the alternative trading institutions in a modeling framework where a random number of informed and liquidity traders, each buying or selling a single unit, chooses between a DM and a CN. In their model, dealers receive multiple single--unit orders and cannot distinguish between the informed and the liquidity orders. Hence, their bid--ask spread corresponds to each order's market impact. \cite{DDH} consider the allocation of order flow between a CN and a DM when trading in both markets takes place at exogenously given prices. They show that small differences in the traders' preferences generate a unique equilibrium, in which patient traders use the CN whereas impatient traders submit orders directly to the DM. Due to the fact that prices are exogenous, the equilibrium market share of the CN is fully determined by the price differential between the markets, together with the distribution of the traders' liquidity preferences. In contrast with the two preceding works, where interactions between DMs and CNs are studied, \cite{Buti} take an alternative approach and analyze a dynamic model with single--unit traders who may place market or limit orders in a limit--order book (LOB). Alternatively, should they have access to it, the agents may place an immediate--or--cancel order in a dark pool (DP). Agents differ in their valuation of the asset and their access to the DP. The authors find that, whenever the LOB is illiquid, the presence of a DP leads to widening spreads and to a decline in the book's depth; thus, to a deterioration of market quality and welfare. This, in spite of the fact that, on average, trade volume increases. These negative effects
are generally decreasing in the depth of the LOB. The take--home message offered is that, when studying interacting LOBs and DPs,
there is a trade--off between trade and volume creation on the one hand, and book depth and spread on the other one.

\vspace{0.2cm}
In terms of the aforementioned effects of the presence of the CN, whereas increases in the number of participating agents and welfare are generic, the narrowing of the spread does not seem to be so. For instance in \cite{Buti}, the presence of a DP results in a migration of liquidity and hence an increasing spread --- an effect that cannot appear in our setting where all traders are liquidity takers. Contrastingly, \cite{Buti_Data}, provide empirical evidence that high DP activity is associated with narrower spreads, but no causality is concluded. In~\cite{Zhu2014}, asymmetric information divides agents into informed and (uninformed) liquidity traders. When a CN complements an existing DM, the spread widens because the liquidity traders move to the CN, whereas the informed ones, who tend to be on one side of the market, prefer the DM. In our setting, agent heterogeneity corresponds to different endowments or preferences, but there is no distinction at the level of access to information. Hence, the spread originates due to the adverse--selection problem faced by the dealer.

\vspace{0.2cm}
The remainder of this article is structured as follows. Our model and main results are presented in Section \ref{sec:Model}. Existence of a solution to the dealer's optimization problem is established in Section \ref{sec:ExistenceSol}.  Section \ref{sec:ImpactSpread} studies the impact of a CN on the spread. Section \ref{sec:ExistenceEqui} establishes our result regarding the existence of equilibrium price schedules. A specific application to a portfolio--liquidation problem with dark--pool trading is analyzed in Section \ref{sec:DPtrading} and Section~\ref{sec:Conclusions} concludes.

\section{\large{Model and main results}}\label{sec:Model}

\noindent We consider a quote--driven market for an asset, in which a risk--neutral \textit{dealer} engages a group of privately--informed \textit{traders}\footnote{Our dealer is called the {\sc principal} in the contract--theory jargon and the traders are usually referred to as the {\sc agents}.}. The dealer market (DM for short) is described by a pricing schedule $T:\re\to\re.$ In other words, $q$ units of the asset are offered to be traded, on a take--it--or--leave--it basis, for the amount $T(q)$. For $q\in\re,$ we refer to the pair $\big(q, T(q)\big)$ as a \textit{contract}. We assume that $T(0)=0$ and that $T$ is absolutely continuous. Thus, we may write
\be
T(q) = \int_0^q t(s)ds,\quad q\geq 0,
\ee
and analogously for negative values of $q.$ Here $t(s)$ is the marginal price at which the $s$--th unit is traded. As we shall see below, pricing schedules are, in general, not differentiable at zero. Hence, for a particular schedule $T$ the \textit{spread} is
\be
\mathcal{S}(T) := |T'(0_+) - T'(0_-)|=|t(0_+) - t(0_-)|,
\ee
where $t(0_-)$ and $t(0_+)$ are the \textit{best--bid} and \textit{best--ask} prices, respectively. We denote by $C:\re\to\re$ the dealer's inventory or risk costs associated with a position $q$, e.g. the impact costs of unwinding a portfolio of size $q$ in a limit order book. We assume that the mapping $q\mapsto C(q)$ is strictly convex, coercive and that it satisfies $C(0) = 0.$

\vspace{0.2cm}
The traders' idiosyncratic characteristics are represented by the index $\t$ that runs over a closed interval $\T:=[\underline{\t}, \overline{\t}],$ called the set of \textit{types}. We assume that zero belongs to the interior of $\T.$ Saying that a trader's type is $\t$ means that if he trades $q$ shares for $T(q)$ dollars his utility is $u(\t, q) - T(q),$ where
\be
u(\t, q):=\t \psi_1(q) + \psi_2(q)
\ee
and $\psi_1,\psi_2:\re\to\re$ are smooth functions that satisfy $\psi_1(0)=\psi_2(0)=0,$  $\psi_1$ is strictly increasing and $C(q)-\Psi_2(q)\geq 0$ holds for all $q\in\re$.
Thus far, with our choice of preferences the traders enjoy a type--independent \textit{reservation utility} of zero, should they decide to abstain from trading in the DM. Such an action is commonly referred to agents choosing their \textit{outside option}. As $C(0) = 0$, providing $\big(0, T(0)\big)$ is costless to the dealer and, since $\big(0, T(0)\big)$ yields all agents their reservation utility, in the absence of any other trading opportunity, we may equate the contract $(0,0)$ to the traders' outside option.

\vspace{0.2cm}
Besides participating in the DM, each trader has the possibility to submit an order to a \textit{crossing network} (CN). The latter is an alternative trading venue where trades take place at fixed bid/ask prices $\pi:=\big(\pi_-, \pi_+\big)$, but where execution might not be guaranteed.\footnote{In other words, the crossing network presents agents with possibly better prices at the cost of an uncertain execution. CN trading often benefits agents who intend to unwind large positions, which might result in a price impact.} The possibility of trading in the crossing network modifies the traders' outside option to the extent that now they may choose between abstaining from all trading and earning zero or participating in the CN if the corresponding expected utility is non--negative. For a specific $\pi,$ the quantity $u_0(\t; \pi)\geq 0$ represents the expected utility of the $\t$--type investor who decides to take his (now extended) outside option. In the sequel we indulge in a slight abuse of the language and also refer to $u_0(\cdot; \pi)$ as the agents' outside option(s). Following \cite{DDH,HM} we focus on the case where a trader chooses exclusively between his outside option and trading in the DM, i.e. we do not allow for simultaneous participation in the DM and the CN.  Initially we take $\pi$ as given, but later we analyze the case where it is endogenously determined through the interaction between the DM and the CN via the feedback of the spread in the former into the pricing in the latter. We work under the following assumption:\footnote{Once an assumption has been made, we consider it to be standing for the remainder of the paper.}

\vspace{0.2cm}
\begin{Assumption}\label{ass:cost of access}
There is a fixed cost $\kappa>0$ of accessing the outside option such that, for all $\pi\in\re^2,$ the function $u_0(\cdot; \pi)$ can be written as
$
u_0(\cdot; \pi) =\max\big\{\widetilde{u}_0(\cdot; \pi) - \kappa, 0\big\},
$
where $\widetilde{u}_0(0; \pi) = 0.$
\end{Assumption}

\vspace{0.2cm}
Trading over the DM is anonymous; the dealer is unable to determine a trader's type before he engages the latter. The only ex--ante information the dealer has is the distribution of the individual types over $\T,$ which is described by a density $f:\T\to\re_+.$ In the sequel we specify the traders' and the dealer's optimization problems and analyze the impact of the CN on the DM, especially on its spread.

\vspace{0.2cm}
\subsection{\large{The traders'  problem}}\label{ssec:Agents}

Until further notice we consider $\pi$ to be fixed. The problem of a trader of type $\t$ is to determine, for a given pricing schedule $T,$
\be
q_m(\t) := \text{argmax}\Big\{u(\t, q) - T(q)\Big\}
\ee
and then choose, for $q_m\in q_m(\t),$  between his \textit{indirect--utility} $v(\t):=u\big(\t, q_m\big) - T\big(q_m\big)$ from trading in the DM and his outside option $u_0(\t; \pi).$ As the supremum of affine functions, the indirect utility function is convex.

\vspace{0.2cm}
The choice of a pricing schedule $T$ induces a partition of the type space. We say that a trader of type $\t$ \textit{participates} in the DM if
$
v(\t)\geq u_0(\t; \pi),
$
assuming that ties are broken in the dealer's favor. Conversely, we say that a trader of type $\t$ \textit{is excluded} from trading in the DM if
$
v(\t) < u_0(\t; \pi).
$
For a given schedule $T,$ we denote the set of excluded types by $\T_e(T;\pi).$ Observe that, in the absence of a CN, there is no loss of generality in assuming that all traders participate. We say that a trader of type $\t$ is \textit{fully serviced} if he earns strictly positive profits from interacting with the dealer.

\vspace{0.2cm}
\subsection{\large{The dealer's problem}}\label{ssec:BenchmarkProblemMM}

The \textit{Revelation Principle} (see, e.g. Meyerson~\cite{Meyerson:91}) says that, when studying Nash--equilibrium outcomes in adverse-selection games such as ours, there is no loss of generality in focusing on direct--revelation mechanisms, i.e. those mechanisms where the set of types indexes the contracts. Furthermore, from the \textit{Taxation Principle} (see, e.g. Rochet~\cite{R:85}) there is also no loss of generality in writing $\tau(\t)$ instead of $T(q(\t)),$ where $\tau:\T\to\re$ is an absolutely continuous function. From this point on we shall, therefore, study our principal--agent game through books of the form $\big\{\big(q(\t), \tau(\t)\big),\t\in\T\big\}$ and drop $T$ from the specification of the indirect--utility functions. We also write $\T_e(q,\tau;\pi)$ instead of $\T_e(T;\pi)$ for the set of excluded types.

\vspace{0.2cm}
At the onset, a trader of type $\t$ could misrepresent his type by choosing a contract $\big(q(\widetilde{\t}), \tau(\widetilde{\t})\big),$ with $\widetilde{\t}\neq \t.$ The dealer strives to avoid this situation, since he wants to exploit the information contained in the density of types. This requires that he offers \textit{incentive--compatible} books, i.e. those that satisfy
\be
\max_{\widetilde{\t}\in\T}\big\{u\big(\t, q(\widetilde{\t})\big) - \tau(\widetilde{\t})\big\} = u\big(\t, q(\t)\big) - \tau(\t).
\ee
In the presence of an incentive--compatible book, the contract that yields a trader of type $\t$ his indirect utility is precisely the one the dealer has designed for him.

\vspace{0.2cm}
Since the dealer is risk neutral, his goal is to maximize his expected income from engaging the traders. Taking into account the impact of the CN on the traders' optimal actions, his problem is to devise $(q^*, \tau^*)$ so as to solve the problem
\be
\begin{array}{cc}
  \mathcal{P}(\pi) := & \left\{
        \begin{array}{l}
          \sup_{(q, \tau)} \int_{\T_e^c(q, \tau;\pi)}\Big(\tau(\t) - C\big(q(\t)\big)\Big)f(\t)d\t\\
          \text{s.t.  }\\
          (q(\t), \tau(\t))\in\text{argmax}_{\widetilde{\t}\in\T}\big\{u\big(\t, q(\widetilde{\t})\big)-\tau(\widetilde{\t})\big\},\\
          \tau \text{  is absolutely continuous}.
        \end{array}
      \right.
\end{array}
\ee

Due to the \textit{Envelope Theorem}, if a contract $\big\{(q(\t), \tau(\t)\big),\t\in\T\big\}$ is incentive compatible, then $\psi_1(q(\t))$ belongs to the subdifferential $\partial v(\t)$. Since for almost all $\t\in\T$ it holds that $\partial v(\t)=v'(\t)$ and $\psi_1$ is strictly increasing, we have for almost all $\t \in \T$ that
\begin{equation}\label{eq:QualGrad}
q(\t) = \psi_1^{-1}\big(v'(\t)\big).
\end{equation}
Therefore, starting from a convex indirect--utility function we can recover, for almost all types, the quantities in the incentive--compatible book that generated it. Furthermore, the indirect utility function may be written as
\begin{equation}\label{eq:IndUt}
\begin{split}
v(\t) & = \t \psi_1\big(\psi_1^{-1}\big(v'(\t)\big)\big) + \psi_2\big(\psi_1^{-1}\big(v'(\t)\big)\big) -\tau(\t) \\
& = \t\,v'(\t) + \psi\big(v'(\t)\big) - \tau(\t),
\end{split}
\end{equation}
where $\psi:=\psi_2\circ \psi_1^{-1}.$ It follows from Eqs.~\eqref{eq:QualGrad} and~\eqref{eq:IndUt} that the traders' indirect utility function contains all the information about the quantities and the pricing schedule, which allows us to write $\T_e^c(v;\pi)$ instead of $\T_e^c(q, \tau;\pi).$ In particular, introducing the functions
\be
	\widetilde{K}(q):=C\big(\psi_1^{-1}(q)\big) - \psi_2\big(\psi_1^{-1}(q)\big)
\quad\text{and}\quad
i(\t, v, q):=\t\cdot q - v - \widetilde{K}(q)
\ee
and denoting by $\C$ the cone of all real--valued convex functions over $\T$, we can restate the dealer's problem as
\be
 \mathcal{P}(\pi) =\sup_{v\in\C} \int_{\T_e^c(v;\pi)} i\big(\t,v(\t),v'(\t)\big)f(\t) d\t.
\ee
We prove in Theorem~\ref{thm:Main1} below that, under suitable assumptions, Problem $\mathcal{P}(\pi)$ admits a solution. The latter is, in fact, quasi--unique in the sense that on the set of participating types the solution is indeed unique. However, agents are excluded by offering their types any incentive--compatible, indirect--utility function that lies below $u_0.$ In other words, there is no uniqueness on the set of excluded types. From the agents' point of view there is no ambiguity: they either trade with the specialist or they take their outside option. The non--uniqueness is also a non--issue for the specialist, since it it only appears in subdomains of the type space that he does not access. With this in mind, in the sequel we denote by $v(\cdot;\pi)$ ``the'' solution to Problem $\mathcal{P}(\pi)$.

\vspace{0.2cm}
\begin{Assumption}\label{ass:qc}
The functions $\psi_1, \psi_2$ and $C$ are such that $\widetilde{K}$ is strictly convex, coercive, continuously differentiable and it satisfies $\widetilde{K}'(0)=0.$
\end{Assumption}

\vspace{0.2cm}
Determining the set of types who do participate but who earn zero profits is essential to our analysis, since it is precisely at the \textit{boundary types}
where $t(0_-)$ and $t(0_+)$ are determined. We prove in Lemma~\ref{lemma:tradingSB} that, by virtue of Assumption~\ref{ass:cost of access}, these limits are always well defined. For any $v\in\C$, we shall refer to
\be
\T_0(v):=\big\{\t\in\T\,|\,v(\t)=0\big\}
\ee
as the set of \textit{reserved traders}. Whenever we refer to the reserved set corresponding to the solution $v(\cdot;\pi)$ to $\mathcal{P}(\pi)$ we write $\T_0(\pi).$ We prove in Proposition~\ref{lm:ZeroatZero} that there is no loss of generality in assuming that any feasible $v\in\C$ satisfies $v(0)=0;$ thus, $\T_0(v)\neq\emptyset.$

\vspace{0.2cm}
\begin{remark}\label{rmk:wellpossed}
A well defined spread requires $\T_0(\pi)$ to be a proper interval $[\underline{\t}_0(\pi), \overline{\t}_0(\pi)],$ which will follow from Assumption~\ref{ass:cost of access}, and that there exists $\epsilon>0$ such that $(\underline{\t}_0(\pi)-\epsilon, \underline{\t}_0(\pi))$ and $(\overline{\t}_0(\pi), \overline{\t}_0(\pi)+\epsilon)$ belong to the set of fully--serviced traders. The existence of such an $\epsilon$ is proved in Lemma~\ref{lemma:tradingSB}. Economically, this conditions means that the CN is not beneficial for low--type traders. We shall encounter several instances where the proofs of our results concern conditions on points to the left of $\underline{\t}_0(\pi)$ or to the right of $\overline{\t}_0(\pi)$ that are analogous. So as to streamline the said proofs, whenever we find ourselves in one of these ``either--or'' situations, we deal only with the positive case.
\end{remark}

\vspace{0.2cm}
\noindent We are now ready to state the first main result of this paper, whose proof is given in Section~\ref{sec:ExistenceSol} below.

\vspace{0.2cm}
\begin{theorem} \label{thm:Main1}
Problem $\mathcal{P}(\pi)$ admits a solution, which is unique on the set of participating types.
\end{theorem}

\vspace{0.2cm}
Our second main result concerns the effect of the CN on the spread and the set of participating traders if, disregarding negative expected unwinding costs, the dealer can match the CN.

\vspace{0.2cm}
\begin{Assumption}\label{ass:matching} There exists an incentive compatible book $\big\{(q_c(\t), \tau_c(\t) \big), \t\in\T\big\}$ such that for almost all $\t\in\T$ it holds that
$
u\big(\t, q_c(\t) \big) - \tau_c(\t)= u_0(\t; \pi).
$
\end{Assumption}

\vspace{0.2cm}
Assumption~\ref{ass:matching} implies that $u_0(\cdot; \pi)$ is also a convex function. The case where $u_0(\cdot; \pi)$ is concave is somewhat simpler, since it boils down to exclusion without matching.

\vspace{0.2cm}
\noindent The following theorem analyzes the impact of the CN on the DM and the traders' welfare.

\vspace{0.2cm}
\begin{theorem}
\label{thm:Main2}
For a given price $\pi=(\pi_-, \pi_+)$ let $\mathcal{S}_m$ and $\mathcal{S}_o$ be the spreads with and without the presence of the crossing network and $v_o$ and $v(\cdot;\pi)$ the corresponding indirect--utility functions, respectively. 
In the presence of the crossing network
\begin{enumerate}

\item less types are reserved, i.e. $\T_0(v_o)\supseteq \T_0(\pi).$ Furthermore, the inclusion is strict if there exists $\t\in\T$ such that $u_0(\t;\pi)>v_o(\t);$

\item if the types are uniformly distributed ($f\equiv(\overline{\t}-\underline{\t})^{-1}$) the spread narrows, i.e. $\mathcal{S}_o \geq \mathcal{S}_m;$

\item the typewise welfare increases, i.e. $v_o(\t)\leq v(\t;\pi)$ for all $\t\in\T.$

\end{enumerate}
\end{theorem}

\vspace{0.2cm}
In the sequel we use the subindexes  $``m"$ and $``o"$ to distinguish structures or quantities with and without a CN, respectively.

\vspace{0.2cm}
\subsection{\large{Equilibrium}}\label{ssec:Equilibrium}

It is natural to assume that pricing in the DM has an impact on the pricing schedule $\pi.$ For example, the CN could be a \textit{dark pool}, where trading takes place at the best--bid and best--ask prices of the primary market. We analyze such an example, within a portfolio--liquidation framework, in Section~\ref{sec:DPtrading}. The pecuniary interaction between the DM and the CN, however, is not unidirectional if the dealer anticipates the effect that his choice of book structure has on the CN. Our main focus is the impact of the CN on the spread in the DM.
Specifically, if we denote by $t(0;\pi):=\big(t(0_-;\pi), t(0_+;\pi)\big)$ the best bid--ask prices in the DM for a given CN price schedule $\pi,$ then we call
$\pi^*$ an \textit{equilibrium price} if $\pi^* = t(0;\pi^*).
$

\vspace{0.2cm}
We make the following natural assumption on the impact of $\pi$ on the traders' outside option.

\vspace{0.2cm}
\begin{Assumption}\label{ass:monotone}
Let $\pi_1\le \pi_2,$ where ``$\leq$'' is the lexicographic order in $\re^2,$ then for all $\t\in\T$ it holds that
$
u_0(\t;\pi_1)\geq u_0(\t;\pi_2).
$
Furthermore, we assume that there exists $\big(\underline{\pi}_-, \overline{\pi}_+\big)\in\re^2$ such that $u_0(\cdot;\pi)\leq 0$ for all $(\pi_-,\pi_+)$ such that ${\pi}_-\leq\underline{\pi}_-$ and $\overline{\pi}_+\leq{\pi}_+.$
\end{Assumption}

\vspace{0.2cm}
\noindent The following is our main result on the existence of an equilibrium price.

\vspace{0.2cm}
\begin{theorem}
\label{thm:Main3}
If types are uniformly distributed, then the mapping $\pi\mapsto t(0; \pi)$ has a fixed point.
\end{theorem}

\vspace{0.2cm}
Summarizing, we have that the dealer can correctly anticipate the movements in prices in the CN when he designs the optimal pricing schedule for the DM. Furthermore, the presence of the CN is beneficial in terms of liquidity, market participation and the traders' welfare.

\vspace{0.2cm}
\begin{remark}
The uniformity of the distribution of types in Theorems~\ref{thm:Main2} and~\ref{thm:Main3} can be relaxed, which is something we postpone to the corresponding proofs, where the required notation is introduced.
\end{remark}

\section{\large{Existence of a solution to Problem $\mathcal{P}(\pi)$}}\label{sec:ExistenceSol}

In this section we prove the existence of a solution to the dealer's problem in the presence of a CN. Even though, strictly speaking, this result is a particular case of Theorem 4.4 in~\cite{Page}, for the reader's convenience we present a proof in our simpler setting. Some of the arguments are somewhat standard, but we include them for completeness. The first important result that we require is that the dealer's optimal choices will lead to him never losing money on types that participate.

\vspace{0.2cm}
\begin{proposition}\label{prop:PosProf}
If $(q^*, \tau^*):\T\to\re^2$ is an optimal allocation, then for all participating types it holds that $\tau^*(\t) - C\big(q^*(\t)\big)\geq 0.$
\end{proposition}

\noindent\begin{Proof} Assume the contrary, i.e. that the set
\be
\widetilde{\T}:=\big\{\t\,|\, v(\t;\pi)\ge u_0(\t;\pi), \tau^*(\t)<C\big(q^*(\t)\big)\big\},
\ee
where $v(\t;\pi)=u\big(\t, q^*(\t)\big) - \tau^*(\t)$ has positive measure. Define a new pricing schedule via
\be
\widetilde{\tau}(\t):=\max\big\{\tau^*(\t), C\big(q^*(\t)\big)\big\}.
\ee
The incentives for types in $\widetilde{\T}^c$ do not change, since their prices remain unchanged, whereas prices for others have increased.
Profits corresponding to trading with types in $\widetilde{\T}$ increase to zero. As a consequence the dealer's welfare strictly increases, which violates the optimality of $(q^*, \tau^*).$
\end{Proof}

\vspace{0.2cm}
A consequence of Proposition~\ref{prop:PosProf} is that, together with Assumption~\ref{ass:qc}, it allows us to restrict the feasible set of the dealer's problem to a compact one. We prove this in several steps,

\vspace{0.2cm}
\begin{lemma}\label{lm:ZeroatZero}
If $v:\T\to\re$ is a non--negative, convex function that solves $\mathcal{P},$ then $v(0)=0.$
\end{lemma}

\noindent\begin{Proof} Assume that $v\in\C$ solves $\mathcal{P}$ and $v(0)>0.$ This implies that $\psi_2\big(q(0)\big) - \tau(0) \geq 0.$ Since, from Assumption~\ref{ass:cost of access}, a trader of type $\t=0$ has no access to a profitable outside option, then he participates. From Proposition~\ref{prop:PosProf} it must then hold that $\tau(0)\geq C\big(q(0)\big)$ which in turn implies that $\psi_2\big(q(0)\big)\geq C\big(q(0)\big).$ This relation, however, can only hold for $q(0)=0,$ which implies that $\tau(0)=v(0)=0.$
\end{Proof}

\vspace{0.2cm}
\begin{lemma}\label{lm:BoundedQ}
There exists $\overline{q}\geq 0$ such that if $v$ is feasible, then $|\partial v|\leq\overline{q}.$
\end{lemma}

\noindent\begin{Proof} From Assumption~\ref{ass:qc} and the compactness of $\T$ we have that the mapping
$ q\mapsto i(\t, v, q) $
tends to $-\infty$ as $|q|\to\infty$ uniformly on $\T$ for $v\geq 0.$ From Proposition~\ref{prop:PosProf}  $i\big(\t, v(\t), v'(\t)\big)$ must be non--negative for all participating types, which concludes the proof.
\end{Proof}

\vspace{0.2cm}
From Lemmas~\ref{lm:ZeroatZero} and~\ref{lm:BoundedQ} we have that the quantity $\max_{\t\in\T}\big\{u_0(\t;\pi)\big\} + \overline{q}\|\T\|$ is an upper bound for any feasible choice of $v,$ which yields the following

\vspace{0.2cm}
\begin{corollary}\label{cor:UnifBound}
The feasible set $\mathcal{A}\subset\mathcal{C}$ of Problem $\mathcal{P}$ is uniformly bounded and uniformly equicontinuous.
\end{corollary}

\noindent\begin{Proof} A uniform bound is $\max_{\t\in\T}\big\{u_0(\t;\pi)\big\} + \overline{q}\|\T\|.$ Lemma~\ref{lm:BoundedQ} guarantees that for any $v\in\mathcal{A}$ it holds that $|\partial v|\leq\overline{q}.$ In other words, $\mathcal{A}$ is composed of convex functions whose subdifferentials are uniformly bounded, hence $\mathcal{A}$ is uniformly equicontinuous.
\end{Proof}

\vspace{0.2cm}
Notice that, when it comes to determining quantities and prices for trader types who do participate, Proposition~\ref{prop:PosProf} results in the dealer having to solve the problem
\be
\begin{array}{cc}
  \widetilde{\mathcal{P}}(\pi):=  & \left\{
        \begin{array}{ll}
          \sup_{v\in\mathcal{A}} \int_{\T}\big(i\big(\t, v(\t), v'(\t)\big)\big)_+ f(\t)d\t\\
          \text{s.t.  } v(\t)\geq u_0(\t;\pi)\text{  for all  } \t\in\T.
        \end{array}
      \right.
\end{array}
\ee
The last auxiliary result that we need is the following proposition, whose proof is a direct consequence of Fatou's Lemma, together with Lemmas~\ref{lm:ZeroatZero} and~\ref{lm:BoundedQ}.

\vspace{0.2cm}
\begin{proposition}\label{prop:USC}
The mapping $v\mapsto\int_{\T}\big(i(\t, v(\t), v'(\t))\big)_+ f(\t)d\t$ is upper semi--continuous in $\mathcal{A}$ with respect to uniform convergence.
\end{proposition}

\vspace{0.2cm}
\noindent We are now ready to prove our first main result:

\vspace{0.2cm}

\noindent\textbf{Proof of Theorem \ref{thm:Main1}:} Assume that $\mathcal{A}\bigcap\big\{v\in\C | v(\cdot)\geq u_0(\cdot ; \pi)\big\}$ is non--empty and consider a maximizing sequence $\big\{\widetilde{v}_n\big\}_{n\in\n}$ of Problem $\widetilde{\mathcal{P}}(\pi).$ From Corollary~\ref{cor:UnifBound} we have that, passing to a subsequence if necessary, there exists $\widetilde{v}\in\mathcal{A}$ such that $\widetilde{v}_n\to\widetilde{v}$ uniformly. A direct application of Proposition~\ref{prop:USC} yields that $\widetilde{v}$ is a solution to $\widetilde{\mathcal{P}}(\pi).$ To finalize the proof we must construct from $\widetilde{v}$ a solution to Problem $\mathcal{P}(\pi).$ To this end, let us define the sets
\be
\T_-:=\big\{\t\in\T | i\big(\t, \widetilde{v}(\t), \widetilde{v}'(\t)\big)<0\big\}\quad\text{and}\quad \T_+:=\T_-^c.
\ee
It is well known that if a sequence of convex functions converges uniformly (to a convex function), then there is also uniform convergence of the derivatives wherever they exist, which is almost everywhere. This fact, together with the continuity of the mappings $\t\mapsto\widetilde{v}(\t)$ and  $(\t,v, q)\mapsto i(\t,v, q),$ implies that $\T_-$ is the union of a disjoint set of open intervals:
\be
\T_-=\bigcup_{i=1}^{\infty}(a_i, b_i).
\ee
Define, for each $i\geq 1,$
\be
\widetilde{v}_{a,i}:=\inf\big\{q | q\in\partial \widetilde{v}(a_i)\big\}\quad\text{and}\quad \widetilde{v}_{b,i}:=\sup\big\{q | q\in\partial \widetilde{v}(b_i)\big\}
\ee
and consider the support lines to $\text{graph}\{\widetilde{v}\}$ at $a_i$ and $b_i$ given by
\be
l_i(\t)=\widetilde{v}(a_i)+\widetilde{v}_{a,i}(\t-a_i)\quad\text{and}\quad L_i(\t)=\widetilde{v}(b_i)+\widetilde{v}_{b,i}(\t-b_i),
\ee
respectively. Let $c_i\in(a_i, b_i)$ be,  for each $i\geq 1,$ the unique solution to the equation $l_i(\t)=L_i(\t)$ and define on $(a_i, b_i)=:\T_i$
\be
\begin{array}{cc}
  v_i^*(\t):=  & \left\{
        \begin{array}{ll}
          l_i(\t) & \t\leq c_i;\\
          L_i(\t) & \t> c_i.
        \end{array}
      \right.
\end{array}
\ee
Finally define
\be
\begin{array}{cc}
  v^*(\t):=  & \left\{
        \begin{array}{ll}
          \widetilde{v}(\t) & \t\in\T_+;\\
          v_i^*(\t) & \t\in\T_i, i\in\n,
        \end{array}
      \right.
\end{array}
\ee
then $v^*$ is a solution to Problem $\mathcal{P}(\pi)$ and $\T_e(v^*)=\T_-,$ which concludes the proof.

\begin{remark}\label{rem:QuasiUniqueness}
If the specialist can profitably match all agents' outside option, then the quasi--uniqueness of a solution to Problem $\mathcal{P}(\pi)$ is in fact uniqueness and it follows directly from Assumption~\ref{ass:qc}. Indeed, in such a case
\be
\big(i(\t, v(\t), v'(\t))\big)_+=\big(i(\t, v(\t), v'(\t))\big)
\ee
 nd problem $\widetilde{\mathcal{P}}(\pi)$ is one of maximizing a strictly concave, coercive functional over a convex set that is closed with respect to uniform convergence. In the general case, we construct the quasi--unique solution in Section~\ref{sec:ModelCN}. Assumption~\ref{ass:qc} remains crucial, since it guarantees that the maximization problems through which we define the optimal quantities have unique maximizers.
\end{remark}

\section{\large{The impact of a crossing network}}\label{sec:ImpactSpread}

{In this section we look at the impact that a CN has on the spread, on participation and on the traders' welfare. In order to do so,
we provide a characterization of the solution to Problem $\mathcal{P}(\pi).$ It should be noted that, given the restriction of candidate solutions to $\C,$ we cannot simply make use of the Euler--Lagrange equations to solve the variational problem, since the said equations are only satisfied when the constraints do not bind.}

\vspace{0.2cm}
\subsection{\large{A benchmark without a CN}}\label{sec:ModelBench}

We first analyze the benchmark case where the traders do not have access to a CN.
The corresponding dealer's problem is denoted by $\mathcal{P}_o.$ Recall that all trader types have a zero reservation utility, which the dealer is able to match costlessly by offering the contract $(0, 0).$ The point of making this normalization is to simplify the constraints in the dealer's optimization problem. This will not be possible in the presence of a CN since, even if the dealer were able to match the utility that investors enjoy if they trade in the CN, this would be in general not costless.

\vspace{0.2cm}
We take a Lagrange--multiplier approach to provide a characterization of the solution to Problem $\mathcal{P}_o.$ To this end, let us introduce the following definition:
\be
I[v] := \int_{\T}i\big(\t, v(\t), v'(\t)\big)f(\t)d\t.
\ee
Let $BV_+(\T)$ be the space of non--negative functions of bounded variation $\gamma:\T\to\re_+,$ which we place in duality with $C(\T, \re),$ the space of real--valued, continuous functions on $\T,$ via the standard pairing
\be
\langle v, \gamma\rangle :=\int_{\T} v(\t)d\gamma(\t)
\ee
for $v\in C(\T, \re),$ where $d\gamma$ is the distributional derivative of $\gamma.$ Furthermore, it follows from Pontryagin's Maximum Principle and the fact that $f$ is a probability density function that there is no loss of generality in assuming that $\gamma$ is absolutely continuous and that $\gamma(\overline{\t}) = 1.$
The Lagrangian for the dealer's problem is
\begin{equation*}\label{eq:Lagrangian}
\mathcal{L}(v, \gamma) := I[v]+\langle v, \gamma\rangle,\quad v\in\C,
\end{equation*}
with corresponding Karush--Kuhn--Tucker conditions
\begin{equation}\label{eq:KT}
\langle v, \gamma\rangle = 0\quad\text{and}\quad d\gamma(\t)=0\Rightarrow v(\t)>0.
\end{equation}
The next result is the formalization of the \textit{vox populi} saying that ``quality does not jump''. Regularity properties of the solutions to variational problems subject to convexity constraints were studied by Carlier and Lachand--Robert in~\cite{CarLach}, and their methodology can be directly adapted to prove the following result.

\vspace{0.2cm}
\begin{proposition}\label{prop:NoJumps1}
If $v\in\C$ is a stationary point of $\mathcal{L}(v, \gamma),$ then $v\in C^1(\T).$
\end{proposition}

\vspace{0.2cm}
The fact that, at the optimum, the mapping $\t\mapsto v'(\t)$ is continuous, implies that $q$ is also a continuous function of the types.  This will prove to be extremely useful, specially in the presence of a crossing network. If we integrate by parts, then $\mathcal{L}(v, \gamma)$ can be transformed into
\be
\Sigma(q, \gamma):=\int_{\T}\bigg(\Big(\t+\frac{F(\t) -\gamma(\t)}{f(\t)}\Big)\psi_1\big(q(\t)\big) - \widetilde{C}\big(q(\t)\big)\bigg)f(\t)d\t,
\ee
where $q(\t) =\psi_1^{-1}\big(v'(\t)\big),$ as described above, and $\widetilde{C}(q):=C(q)-\Psi_2(q).$ The idea now is to maximize the mapping
\be
q \mapsto \sigma(\t, q, \Gamma) := \Big(\t +\frac{F(\t) -\Gamma}{f(\t)}\Big)\psi_1(q) - \widetilde{C}\big(q\big)
\ee
pointwise, for a given fixed $\Gamma$ (in the sequel we use $\Gamma$ whenever we are dealing with an arbitrary but fixed value of $\gamma$). From Assumption~\ref{ass:qc} it follows that we can write down the unique maximizer as
\be
l(\t, \Gamma):=K^{-1}\Big(\frac{F(\t) + \t\,f(\t) -\Gamma}{f(\t)}\Big),
\ee
where $K(q):=\widetilde{C}'(q)/\Psi_1'(q).$ For each $\t\in\T$ and $\Gamma\in [0, 1],$ the quantity $l(\t, \Gamma)$ is a candidate for the optimal $q(\t)$ and convexity (or incentive compatibility) is verified if the mapping $\t\mapsto l(\t, \Gamma)$ is increasing. The crux is then to determine the Lagrange multiplier $\gamma.$ In the sequel we denote $\T_o:=\T_0(v_o^*),$ where $v_o^*$ solves Problem $\mathcal{P}_o.$ In other words, if $\t\in\T_o$, then $q(\t)=T(\t)=v(\t)=0.$

\vspace{0.2cm}
From Lemma~\ref{lm:ZeroatZero} we have that, unless $v(\underline{\t})=0,$ the quantity $q(\underline{\t})<0$ and the complementary--slackness condition imply that $\gamma(\t)=0$ for $\t\in[\underline{\t}, \widetilde{\t})$ for some $\widetilde{\t}>\underline{\t}.$ The left endpoint $\underline{\t}_0$ of $\T_o$ is then determined by solving the equation
\be
K^{-1}\Big(\t + \frac{F(\t)}{f(\t)}\Big)=0.
\ee
Furthermore, since $v$ must be convex, once $v(\hat{\t})>0$ then $v(\t)>0$ for all $\t>\hat{\t}.$ This implies that the right endpoint $\overline{\t}_0$ of $\T_o$ is determined by solving the equation
\be
K^{-1}\Big(\t-\frac{1 - F(\t)}{f(\t)}\Big)=0.
\ee
The quantities $F(\t)/f(\t)$ and  $(1 - F(\t))/f(\t)$ are know as the \textit{hazard rates}, and sufficient conditions for the mapping $\t\mapsto l(\t, \Gamma)$ to be non--decreasing are
\be
\frac{d}{d\t}\left(\frac{F(\t)}{f(\t)}\right)\geq 0 \geq \frac{d}{d\t}\Big(\frac{1 - F(\t)}{f(\t)}\Big),
\ee
see, e.g. Biais et al.~\cite{BMR} for a discussion on this condition.

\vspace{0.2cm} Let us assume that we have determined $\T_o$. What remains is then to connect the participation constraint with the spread. Differentiating Eq.~\eqref{eq:IndUt} and noting that $v'(\t)=\psi_1(q(\t))$ we have that
\be
\tau'(\t) = q'(\t)\big(\t\psi_1'(q(\t)) + \psi_2'(q(\t))\big).
\ee
Observe that $\tau'(\underline{\t}_0)$ and $\tau'(\overline{\t}_0)$ are in fact $T'(0_-)$ and $T'(0_+),$ since by construction $q(\underline{\t}_0) = q(\overline{\t}_0)=0.$ If we define
$\phi_1:= \psi_1'(0)$ and $\phi_2:=\psi_2'(0)$, then we have that the spread is given by the expressions
\begin{equation}\label{eq:Spread}
t(0_-) = q'(\underline{\t}_{0}-)\big(\underline{\t}_0\phi_1 + \phi_2\big)\quad\text{and}\quad t(0_+) = q'(\overline{\t}_{0}+)\big(\overline{\t}_0\phi_1 + \phi_2\big).
\end{equation}
Our objective in Section~\ref{sec:ModelCN} is to compare the values above to those obtained in the presence of a crossing network.

\vspace{0.2cm}
Before we proceed we present two examples so as to illustrate the use of the methodology described hitherto. The first revisits Mussa \& Rosen \cite{MR:78}. The second is slightly more advanced. We shall use it below to illustrate the complex structure of equilibrium pricing schedules and utilities in the presence of CNs.

\vspace{0.2cm}
\begin{example} \label{MussaRosen} Let us assume that $\T = [-r,r]$ for some $r>0$, that types are uniformly distributed and that
\be
	u(\t,q) = \t q.
\ee
We also set $C(q)=0.5\,q^2.$ By direct computation we find that $\underline{\t}_0 = -\frac{r}{2}$ and $\overline{\t}_0 = \frac{r}{2}.$
Since a trader of type $\t \in \T_o$ is brought down to reservation utility and hence trades $q(\t)=0,$ the expression
\be
q(\t) = \t + \frac{F(\t) - \gamma(\t)}{f(\t)}=2\t+r-2r\gamma(\t)
\ee
implies that the Lagrange multiplier is
\be
	\gamma(\t) = \left\{ \begin{array}{ll} 	0 & \t < \underline{\t}_0 \\
											\frac{1}{2} + \frac{\t}{r} & \t \in \T_o \\
											1 & \t > \overline{\t}_0
											\end{array} \right. .
\ee
In particular, $q'(\underline{\t}_{0}-) = q'(\overline{\t}_{0}+) = 2$ and hence $t(0_-) = -r$ and $t(0_+) = r$. Thus, the spread increases linearly in the highest/lowest type.
\end{example}

\vspace{0.2cm}
\begin{example} \label{ex1}
Let us assume that the distribution of types over $\T=[-1,1]$ is given by $f(\t)=(2\t+3)/4$ for $\t \in [-1,0)$ and $f(\t)=(3-2\t)/4$ for $\t \in [0,1]$; that $C(q)=0.5\,q^2$ and that $u(\t, q)-\tau = \t\cdot q + 0.25\,q^2-\tau.$ It is straightforward to show that the conditions on the Hazard rates are satisfied and that
\be
K^{-1}\Big(\t+\frac{F(\t)}{f(\t)}\Big)={2}\Big[\frac{3\t^2+6\t+2}{2\t+3}\Big]
\quad\text{and}\quad
K^{-1}\Big(\t-\frac{1 - F(\t)}{f(\t)}\Big)={2}\Big[\frac{3\t^2-6\t+2}{2\t-3}\Big].
\ee
Furthermore, $\T_o\approx\big[-0.423, 0.423\big].$ For the spread, we have that $t(0_-)=q'(\underline{\t}_0)\underline{\t}_0\approx -1.359$ and $t(0_+)=q'(\overline{\t}_0)\overline{\t}_0\approx 1.359.$ In order to obtain $v$ we integrate $q$ (since $\Psi_1(q)=q$) and take into account that $v\equiv 0$ over $\T_o.$ We plot graph$\{v_o\}$ in Figure~\ref{fig:Benchmark}, as well as the per--type profits of the dealer.

\begin{figure}[ht!]
\begin{tabular}{cc}
\hspace{-0.5cm}\subfigure[\label{fig:IndUt} Indirect Utilities]
{\includegraphics[width=0.52\textwidth]{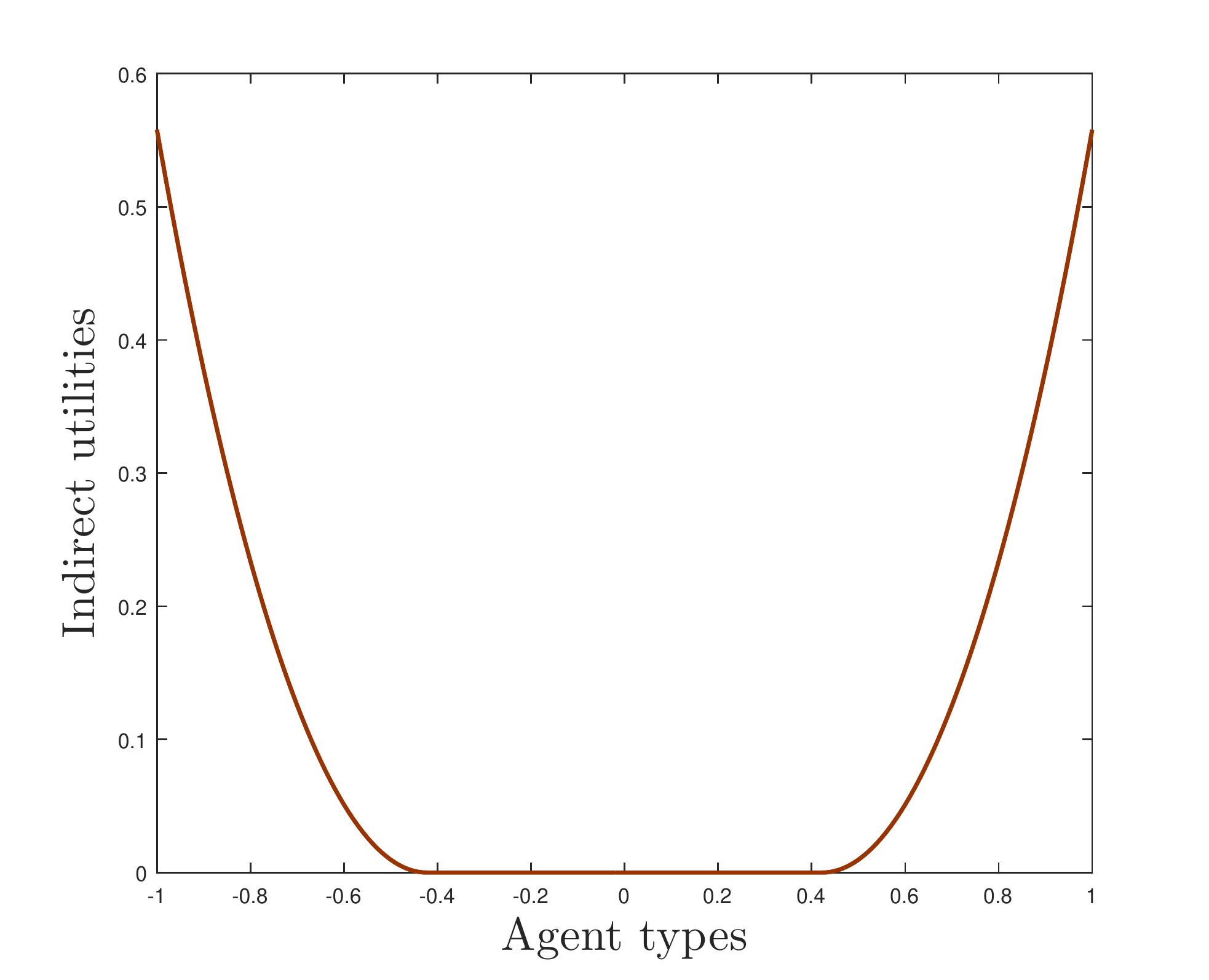}} &
\subfigure[\label{fig:SpecProf}Profits]
{\includegraphics[width=0.52\textwidth]{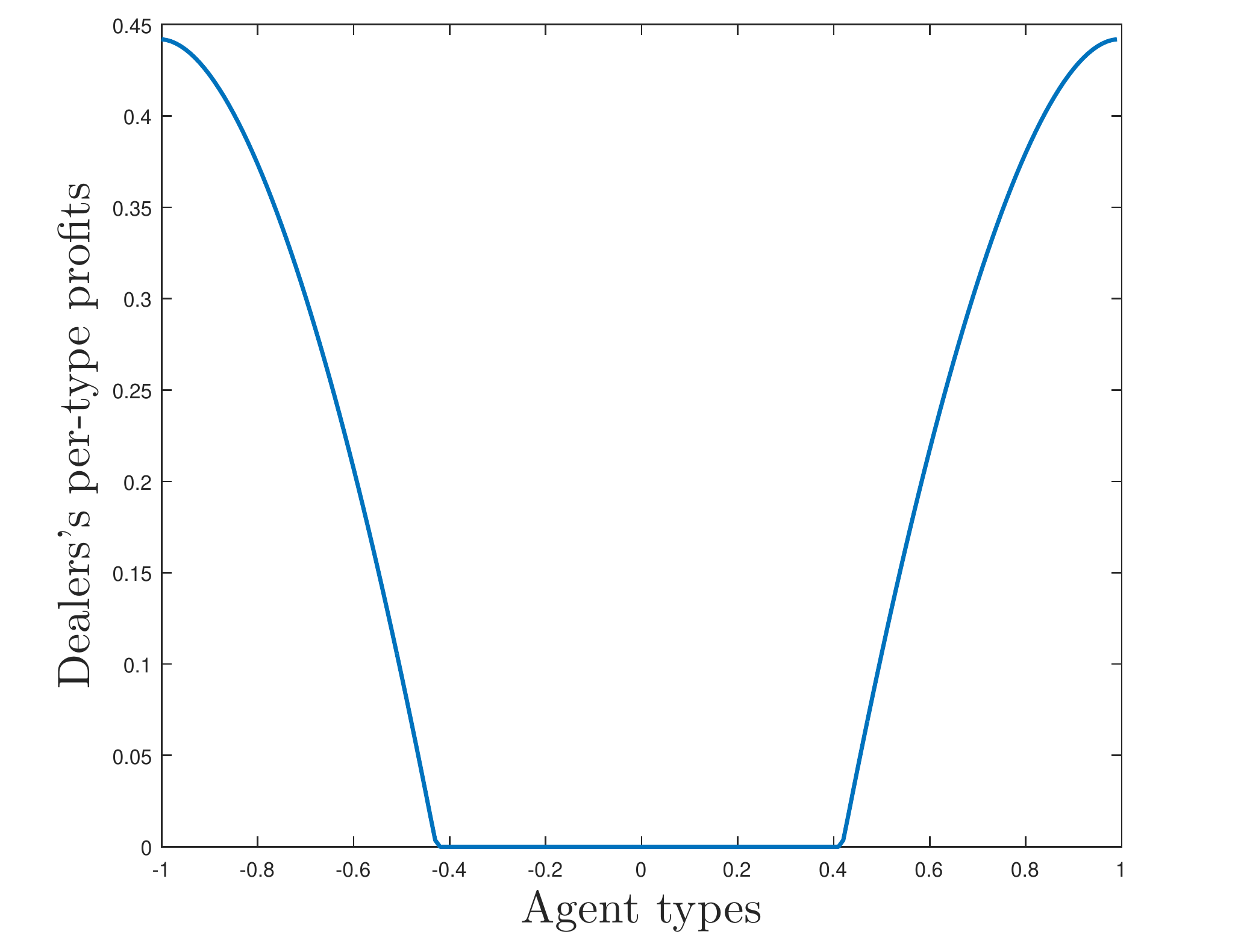}} \\
\end{tabular}\caption{An example without a crossing network}\label{fig:Benchmark}
\end{figure}
\end{example}

\subsection{\large{Introducing a crossing network}}\label{sec:ModelCN}

Let us now analyze the dealer's problem when the market participants have access to a CN that yields a trader of type $\t$ the expected utility $u_0(\t; \pi).$ In this setting it is no longer without loss of generality to assume that all traders participate in the DM, given that enforcing participation (which can be done thanks to Assumption~\ref{ass:matching}) may result in losses to the dealer. The latter may, as a consequence, choose to abstain from trading with a set of types $\T_e(v)$ by offering an incentive--compatible book whose corresponding indirect--utility function lies strictly under $u_0(\t; \pi)$ for $\t\in\T_e(v)$. The resulting problem for the dealer would be
\be
 \mathcal{P}(\pi) =  \sup_{v\in\C} \int_{\T}\Big(\t\,v'(t)  - v(t) - \widetilde{K}\big(v'(\t)\big)\Big)\i_{\{\T_e^c(v)\}}(\t)f(\t)d\t.
\ee
Dealing with the presence of the zero--one indicator function $\i_{\{\T_e^c\}}$ is quite cumbersome (see, e.g. Horst \& Moreno--Bromberg~\cite{HoMo2}), since its domain of definition may change with different book choices. In contrast to the setting studied in~\cite{HoMo2}, however, here the CN is passive. This lack of non--cooperative--games component allows for an alternative way to proceed. To this end, we make use of the following \textit{accounting trick}, which was introduced by Jullien~\cite{BJ:03}: Let us assume that the dealer had access to a fictitious market such that the unwinding costs from trading in it, denoted in the sequel by $C_c,$ satisfy $C_c(q(\t)) = \tau(\t)$ for almost all $\t\in\T.$ In this way, we may again assume that the dealer trades with all market participants, but now his costs of unwinding are given by the function $\mathbb{C}:\re\to\re$ defined as
\be
\mathbb{C}(q):=\min\big\{C(q), C_c(q)\big\},\quad q\in\re.
\ee
In terms of incentives, nothing is distorted by introducing the cost function $\mathbb{C},$ but we must identify the points where there is switching from using $C$ to using $C_c$ and vice versa. These switching points will determine the regions of market segmentation.

\vspace{0.2cm}
If we define, for any traded quantity $q,$ the function $\widetilde{\mathbb{C}}(q) := \mathbb{C}\big(q\big) - \psi_2\big(q\big),$ then we may re--use the machinery from Section~\ref{sec:ModelBench} with minor modifications;\footnote{Observe that Assumptions~\ref{ass:cost of access} and~\ref{ass:matching} imply that $\widetilde{\mathbb{C}}$ satisfies Assumption~\ref{ass:qc}.} namely, denoting by $\mathbb{I}$ the energy corresponding to the cost function $\mathbb{C},$ we may write the Lagrangian of the dealer's problem as
\begin{equation*}\label{eq:LagrangianCN}
\mathbb{L}(v, \gamma) := \mathbb{I}[v]+\langle v - u_0(\cdot; \pi), \gamma\rangle,
\end{equation*}
with the corresponding complementary--slackness conditions. From here we may proceed as in Section~\ref{sec:ModelBench} to find the quantities that the dealer will choose to offer. Strictly speaking we should find the pointwise maximizer in $q$ of the expression
\begin{equation}\label{eq:VirtualSurplus}
\Big(\t  +\frac{F(\t) -\Gamma}{f(\t)}\Big)\Psi_1(q) - \mathbb{K}(q),
\end{equation}
where $\mathbb{K}(q):=\widetilde{\mathbb{C}}(q)-\Psi_2(q).$ This may fortunately be avoided, given that whenever $\mathbb{C}(q)=C_c(q)$, the participation constraint binds and $q(\t)=q_c(\t).$ Before proceeding to the proof of Theorem~\ref{thm:Main2}, we study the mechanism used by the dealer to choose between excluding types, matching the CN and trading with them while offering strictly positive rents.

\vspace{0.2cm}
Whenever the participation constraint does not bind, the dealer selects the quantity to be chosen via the pointwise maximization of the mapping $q\mapsto \sigma(\t, q, \Gamma).$ What makes the current problem trickier than the case without a CN is that now we must pay more attention to the evolution of the multiplier $\gamma.$ If we compare $l(\t, 0)$ and $l(\t, 1)$ to $q_c(\t)$ we may pinpoint the set where the participation constraint may bind. Observe that $\big\{l(\t, 1),\t\in\T\big\}$ and $\big\{l(\t, 0), \t\in\T\big\}$ are the sets of the lowest and highest quantities the dealer may offer in an individually--rational way. Hence, as long as $l(\t, 1)\leq q_c(\t) \leq l(\t, 0)$ there is the possibility of \textit{profitable matching}.

\vspace{0.2cm}
There might be instances where the participation constraint  is binding for some type $\t\in\T$, i.e. $\big(q(\t), \tau(\t)\big)=\big(q_c(\t), \tau_c(\t)\big),$ and $\tau_c(\t) - C\big(q_c(\t)\big)<0.$ In such cases $\mathbb{C}\big(q_c(\t)\big)=C_c\big(q_c(\t)\big)$  and $\t\in\T_e(v)$ for the corresponding indirect utility function, and we say there is \textit{exclusion}.

\vspace{0.2cm}
\begin{remark}\label{rem:QuasiUniquenessReprise}
It is at this point that the quasi--uniqueness mentioned in Remark~\ref{rem:QuasiUniqueness} can be addressed. The principal's problem $\mathbb{P}(\pi)$ using the cost function $\mathbb{C}$ results in the condition
\be
\big(i(\t, v(\t), v'(\t))\big)_+=\big(i(\t, v(\t), v'(\t))\big)
\ee
 being trivially satisfied. As a consequence,
problem $\mathbb{P}(\pi)$ admits a unique solution. The latter coincides, by construction, with the solution to $\mathcal{P}(\pi)$ whenever $\mathbb{C}(q(\t))=C(q(\t))$. The caveat is that the solution to problem $\mathbb{P}(\pi)$ is blind towards what is offered to excluded types, since here their outside option is costlessly matched (they are effectively reserved). Constructing incentive compatible contracts for the excluded types is, thanks to the convexity of the indirect utility function, relatively simple. For instance if an interval of types $(\t_1, \t_2)$ were excluded (but $\t_1$ and $\t_2$ participated) one could consider any two supporting lines to $graph\{v(\cdot;\pi)$ at $(\t_1, v(\t_1;\pi))$ and $\t_2, v(\t_2;\pi)$. From the resulting indirect--utility function on $(\t_1, \t_2)$ one could extract the corresponding quantities and prices. The resulting global convexity of the indirect--utility function offered by the principal would imply that all incentives would remain unchanged. Whether the principal would suffer losses from the contracts offered to types on $(\t_1, \t_2)$ would be irrelevant, since the corresponding agents do not participate.
\end{remark}

\vspace{0.2cm}
As mentioned above, here it is not necessary to determine $\gamma(\t)$ in order to do likewise with $q(\t).$ On the other hand, however, if we interpret $\gamma$ as the shadow cost of satisfying the participation constraint, we may wish to identify the multiplier so as to have a measure of the impact of the CN on the dealer's profits. The following result, which deals with points where there is switching between matching and fully servicing, extends Proposition~\ref{prop:NoJumps1}.

\vspace{0.2cm}
\begin{proposition}\label{prop:NoJumps2} For $\pi\in\re^2$ given, let $\widetilde{\t}\in\T$ be such that there exists $\epsilon>0$ such that $v(\t;\pi)=u_0(\t;\pi)$ on $(\widetilde{\t}-\epsilon, \widetilde{\t}]$ and $v(\t;\pi)>u_0(\t;\pi)$ on $(\widetilde{\t}, \widetilde{\t}+\epsilon]$. Furthermore, assume that
\be
\int_{\widetilde{\t}-\epsilon}^{\widetilde{\t}}\big(\tau(\t)-C(q(\t))\big)f(\t)d\t>0,
\ee
where $\big\{\big( q(\t), \tau(\t) \big), \t \in \T \big\}$ implements $v(\cdot;\pi).$ In other words, there is profitable matching on $(\widetilde{\t}-\epsilon, \widetilde{\t}]$ and the dealer fully services types on $(\widetilde{\t}, \widetilde{\t}+\epsilon].$ Then $\partial v(\widetilde{\t};\pi)$ is a singleton. The result also holds if the order of the matching and full-servicing intervals
is switched.
\end{proposition}

\vspace{0.2cm}
\noindent The rationale behind Proposition~\ref{prop:NoJumps2} is that, as long as the dealer is able to match the traders' outside option without incurring in a loss, it is possible to normalize the latter to zero and directly apply Proposition~\ref{prop:NoJumps1}. This is, naturally, not the case when matching $u_0$ results in losses. We put Proposition~\ref{prop:NoJumps2} to work in Example~\ref{RichStructure}.

\vspace{0.2cm}
Before moving on, we present below a modification to Example~\ref{ex1} that shows how even agents without access to a non--trivial outside option benefit from the presence of the CN and that the optimal Lagrange multiplier need not be continuous.

\vspace{0.2cm}
\begin{example}\label{ex2}


Let $f,$ $\T,$ $C$ and $u$ be as in Example \ref{ex1} and assume that the CN offers the traders the following expected profits:
\be
u_0(\t; (3.2, 3.2))=\begin{cases}
					-0.975\t - 0.52, \text{ if } \t\leq -\frac{8}{15};\\
					0.975\t - 0.52, \text{ if } \t\geq \frac{8}{15};\\
					\text{convex and negative for } \t\in(-\frac{8}{15}, \frac{8}{15}).
					\end{cases}
\ee
Matching this outside option would require the dealer to offer the contracts $(\pm 0.975, 0.52)$. This is profitable, hence the indirect utility never lies below $u_0$. To illustrate this, we have plotted the indirect--utility function in Figure~\ref{fig:IndUtNoEx}. It strictly dominates the one plotted in Figure~\ref{fig:IndUt} for all types who earn positive profits. The smooth pasting condition ($l(\t,\gamma(\t))=q_c(\t)$ where $v$ touches $u_0$, i.e. in $\pm 0.675$) determines the optimal Lagrange multiplier, namely $\gamma(-1)=0$ and $\gamma \equiv 0.030$ on $(-1,-0.389]$. For positive types we obtain symmetrically $\gamma(1)=1$ and $\gamma \equiv 0.970$ on $[0.389,1)$.
The new spread, given by $\big(t(0_-), t(0_+)\big)=(-1.282, 1.282)$, is strictly smaller than in the case without a CN.

\begin{figure}[ht!]
\begin{tabular}{cc}
\hspace{-0.5cm}\subfigure[\label{fig:IndUtNoEx} Indirect Utilities]
{\includegraphics[width=0.52\textwidth]{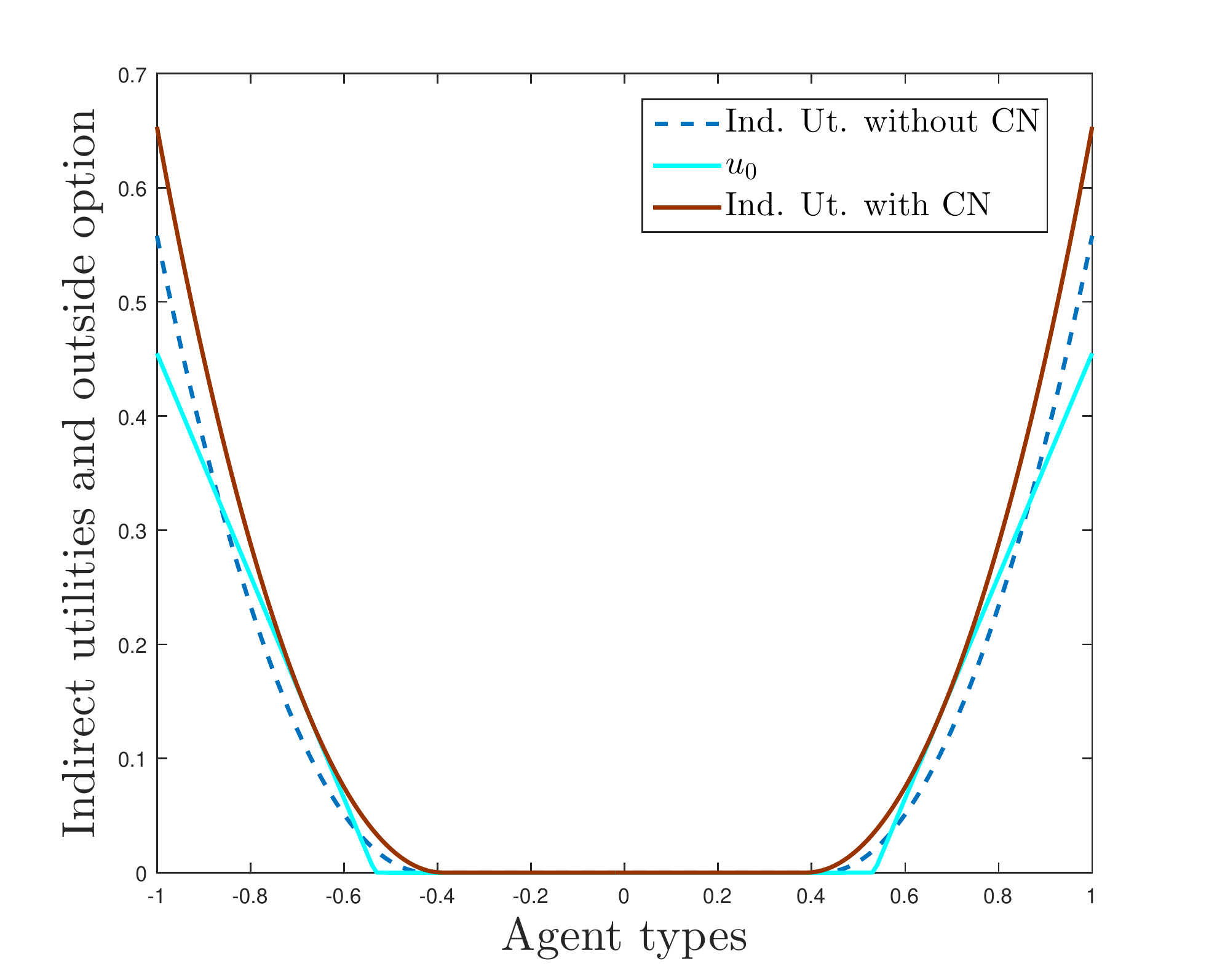}} &
\subfigure[\label{fig:LMNoEx} Lagrange Multiplier]
{\includegraphics[width=0.52\textwidth]{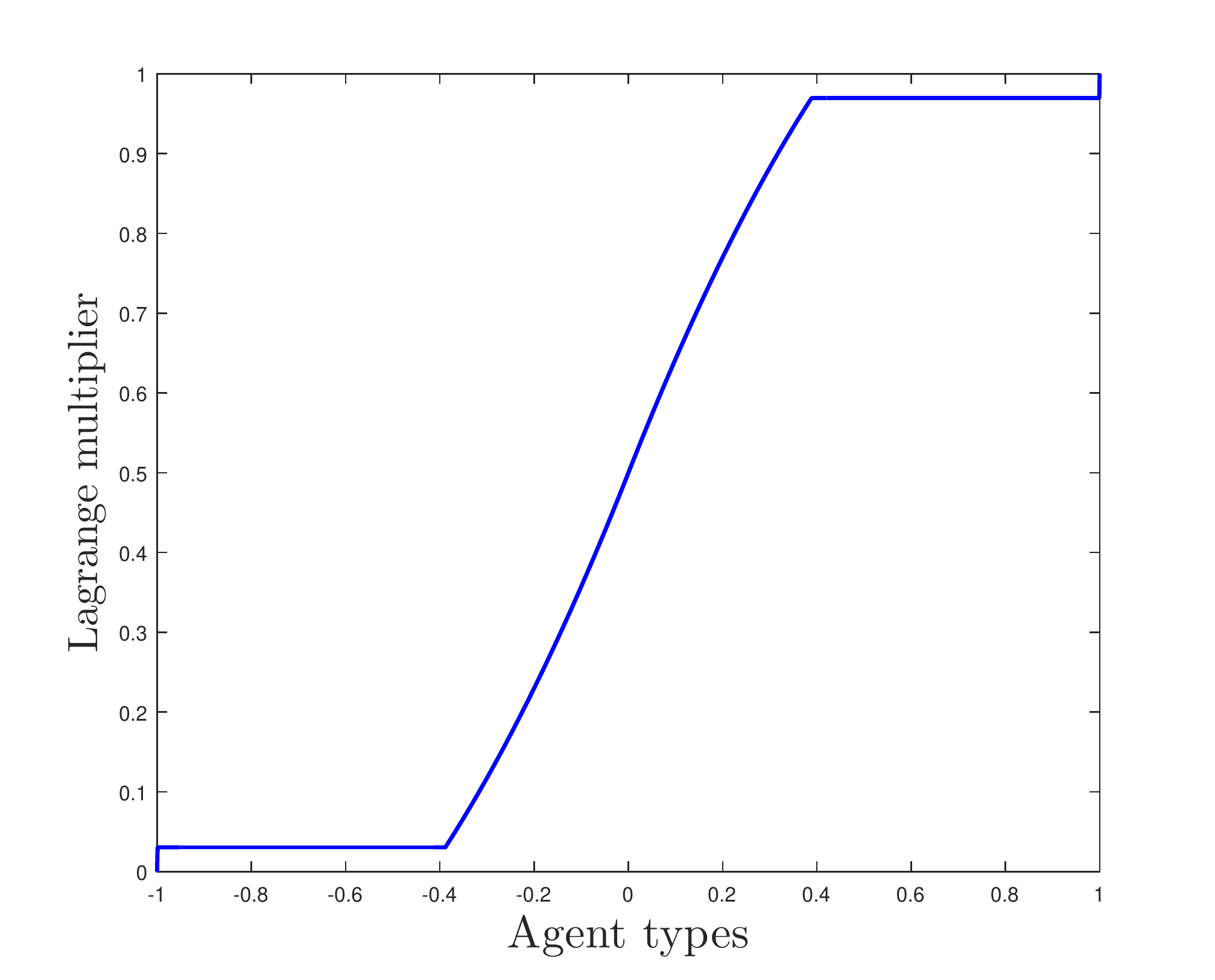}} \\
\end{tabular}\label{fig:NoExclusion}\caption{An example without exclusion}
\end{figure}
\end{example}

The following result will prove to be essential for the results in Section~\ref{sec:ExistenceEqui}. It guarantees,by virtue of Assumption~\ref{ass:cost of access}, our notion of the spread is well defined in the presence of a CN and could be loosely summarized by saying that the first (in terms of moving away from $\t=0$) types to earn positive utility trade in the DM.

\vspace{0.2cm}
\begin{lemma}\label{lemma:tradingSB} There exists $\epsilon=\epsilon(\pi)$ such that the types that belong to $(\underline{\t}_0(\pi)-\epsilon,\underline{\t}_0(\pi) )\cup (\overline{\t}_0(\pi), \overline{\t}_0(\pi)+\epsilon)$ are fully serviced.
\end{lemma}

\noindent\begin{Proof} Let us denote by $\hat{\t}$ the positive solution to the equation $u_0(\t;\pi)=0.$ If there exists $\eta>0$ such that types on $(\hat{\t}, \hat{\t}+\eta)$ can be matched profitably, then the result follows either because $\overline{\t}_0(\pi)<\hat{\t}$ or because $\overline{\t}_0(\pi)=\hat{\t}$ and
the types on $(\hat{\t}, \hat{\t}+\epsilon),$ for some $0<\epsilon\leq \eta,$ are fully serviced. Let us now assume that such an $\eta$ does not exist, we claim then that
$\overline{\t}_0(\pi)<\hat{\t}$ must hold. Proceeding by the way of contradiction, let us assume that $\overline{\t}_0(\pi)=\hat{\t}$ (which is equivalent to $\overline{\t}_0(\pi)\geq\hat{\t}$) and that there exists $\delta>0$ such that $(\hat{\t}, \hat{\t}+\delta)\subset\T_e(\pi).$ This configuration can be improved upon as follows: let $a>0$ be such that $\hat{\t}-a>0.$ By construction $l(\hat{\t}-a, \gamma(\hat{\t}-a))=0.$ Let us fix $\gamma(\t)\equiv \gamma(\hat{\t}-a)=:\Gamma(a)$ for $\t\in(\hat{\t}-a, \t_a),$ where $\t_a$ the solution to $v_a(\t)=u_0(\t;\pi)$ on $(\hat{\t}-a,\overline{\t}]$ if it exists or $\t_a=\overline{\t}$ otherwise, given that we denote by $v_a$ the indirect--utility function corresponding to $\Gamma(a)$. In particular $\t_a>\hat{\t}$ and $l(\t,\Gamma(a))>0$ for $\t \in (\hat{\t}-a, \t_a)$.

We now have that types $\t\in(\hat{\t}-a, \t_a)$ are fully serviced. By Assumption~\ref{ass:cost of access}, $v'_a(\hat{\t}-a)=0<u_0'(\hat{\t};\pi);$ therefore,  there exists $a_1>0$ such that for all $a\leq a_1$ it holds that $\t_a < \hat{\t}+\delta.$ If we could show that there exists $a\leq a_1$ such that the principal could  offer types in $(\hat{\t}-a, \t_a)$ the quantities $q_a(\t)=l(\t, \Gamma(a))$ at a profit, we would contradict the optimality of $\overline{\t}_0(\pi)$ and the proof would be finalized, since incentives above $\t_a$ would not be distorted and the principal's profits would strictly increase. In order to do so, observe that the principal's typewise profit when offering  $q_a(\t)$ is
\be
P(\t):=\t\Psi_1(q_a(\t))+\Psi_2(q_a(\t))-v_a(\t)-C\big(q_a(\t)\big).
\ee
In particular, $P(\hat{\t}-a)=0$ and
\begin{align*}
P'(\hat{\t}-a)& =\Psi_1(q_a(\hat{\t}-a))+(\hat{\t}-a)\Psi_1'(q_a(\hat{\t}-a))q_a'(\hat{\t}-a)+v_a'(\hat{\t}-a)-\widetilde{C}'\big(q_a(\hat{\t}-a)\big)q_a'(\hat{\t}-a)\\
							& =\Psi_1(0)+(\hat{\t}-a)\Psi_1'(0)q_a'(\hat{\t}-a)+v_a'(\hat{\t}-a)-\widetilde{C}'\big(0\big)q_a'(\hat{\t}-a)\\
							& = (\hat{\t}-a)\Psi_1'(0)q_a'(\hat{\t}-a).
\end{align*}
The step from the second to the third equality follows, because by construction $v_a'(\hat{\t}-a)=0;$ by assumption $\Psi_1(0)=0$ and, from Assumption~\ref{ass:qc}, $\widetilde{C}'\big(0\big)=0.$ Furthermore, since $\Psi_1$ is strictly increasing and $q_a'(\hat{\t}-a)>0,$ then $P'(\hat{\t}-a)>0.$ Therefore, there exists $b>0$ such that $P(\t)> 0$ if $\t\in(\hat{\t}-a, \hat{\t}-a+b).$ As a consequence, if $a<a_1$ is small enough, then $P(\t)>0$ for $\t\in(\hat{\t}-a, \t_a),$ as required.

\end{Proof}

\vspace{0.2cm}
\noindent We are now in the position to present the proof of our second main result.

\vspace{0.2cm}

\noindent\textbf{{Proof of Theorem~\ref{thm:Main2}.}} 
(1) Observe that if $\pi$ is such that $\big(\underline{\t}_0(\pi), \overline{\t}_0(\pi)\big)=\T_0(\pi)\subset \T_o,$ then the result follows immediately from Lemma~\ref{lemma:tradingSB}. If we revert the inclusion, two situations are possible, since the addition of the CN--constraint to Problem $\mathcal{P}_o$ may or may not bind for some types. The latter case being trivial, let us look at the case where there is a point $\t_a>\overline{\t}_0$ on which it holds that $v_o(\t_a)=u_0(\t_b;\pi)$ and such that $v_o(\t)>u_0(\t;\pi)$ for $\t<\t_a$ and vice versa for $\t>\t_a.$ The Lagrange multiplier $\gamma_m$ is active on $(\t_a, \overline{\t}],$ which implies that $\gamma_m(\t_a)<1.$ We know from~\cite{BJ:03}, p. 9, that for all $\t$ such that $l(\t, \Gamma)>0,$ the latter is decreasing in $\Gamma.$ As a consequence, the root of the equation
\be
K^{-1}\Big(\t+\frac{F(\t)-\gamma_m(\t_a)}{f(\t)}\Big) = 0
\ee
is strictly smaller than that of $l(\t,1)=0,$ which yields the desired result.

\vspace{0.2cm}
\noindent (2) Let us denote by $t_o(0_-)$ and $t_o(0_+)$ the best bid and ask prices without the presence of a CN and by $t_m(0_-)$ and $t_m(0_+)$ the corresponding marginal prices with one; thus,
\be
t_o(0_-)=q_o'(\underline{\t}_{0,o-})\big(\underline{\t}_{0,o}\phi_1 + \phi_2\big)\text{ and } t_o(0_+) = q_o'(\overline{\t}_{0,o+})\big(\overline{\t}_{0,o}\phi_1 + \phi_2\big)
\ee
and
\be
t_m(0_-)=q_m'(\underline{\t}_{0,m-})\big(\underline{\t}_{0,m}\phi_1 + \phi_2\big)\text{ and } t_m(0_+) = q_m'(\overline{\t}_{0,m+})\big(\overline{\t}_{0,m}\phi_1 + \phi_2\big).
\ee
From Part (1) we know that $\underline{\t}_{0,o}\leq\underline{\t}_{0,m}$ (both negative) and $\overline{\t}_{0,m}\leq\overline{\t}_{0,o}$ (both positive) and, since $\phi_1$ and $\phi_2$ do not depend on the presence of the CN, all we have left to do is show that
\be
q_m'(\underline{\t}_{0,m-})\leq q_o'(\underline{\t}_{0,o-})\text{ and }q_m'(\underline{\t}_{0,m+})\leq q_o'(\underline{\t}_{0,o+}).
\ee
Using the well--known relation $(f^{-1})'(a)=1/f'(f^{-1}(a))$ we have that
\begin{align*}
q_m'(\underline{\t}_{0,m-})& = \frac{1}{K'\Big(K^{-1}\big(\underline{\t}_{0,m} - \frac{\gamma(\underline{\t}_{0,m-})-F(\underline{\t}_{0,m})}{f(\underline{\t}_{0,m})}\big)\Big)}\frac{d}{d\t}\big(\t-\frac{\gamma(\t)-F(\t)}{f(\t)}\big)\Big|_{\t=\underline{\t}_{0,m-}}\\
                           & =  \frac{1}{K'\big(q_m(\underline{\t}_{0,m})\big)}\frac{d}{d\t}\big(\t-\frac{\gamma(\t)-F(\t)}{f(\t)}\big)\Big|_{\t=\underline{\t}_{0,m-}}\\
                           & =  \frac{1}{K'(0)}\Big(1-\frac{d}{d\t}\big(\frac{\gamma(\underline{\t}_{0,m-})-F(\t)}{f(\t)}\big)\Big)\Big|_{\t=\underline{\t}_{0,m}},
\end{align*}
where we have used the fact that $\gamma$ is constant on $(\underline{\t}_{0,m}-\delta,\underline{\t}_{0,m})$ for some $\delta>0.$ We may proceed analogously for the other three quantities. We have to show that
\begin{align}\begin{split}\label{eq:condMonot}
\frac{1}{K'(0)}\frac{d}{d\t}\Big(\frac{\gamma(\underline{\t}_{0,m-})-F(\t)}{f(\t)}\Big)\Big|_{\t=\underline{\t}_{0,m}} & \geq\frac{1}{K'(0)}\frac{d}{d\t}\Big(\frac{-F(\t)}{f(\t)}\Big)\Big|_{\t=\underline{\t}_{0,o}}\\
\frac{1}{K'(0)}\frac{d}{d\t}\Big(\frac{\gamma(\overline{\t}_{0,m+})-F(\t)}{f(\t)}\Big)\Big|_{\t=\overline{\t}_{0,m}} & \geq\frac{1}{K'(0)}\frac{d}{d\t}\Big(\frac{1-F(\t)}{f(\t)}\Big)\Big|_{\t=\overline{\t}_{0,o}},
\end{split}
\end{align}
which hold with equality under the assumption that $f\equiv(\overline{\t}-\underline{\t})^{-1}.$

\vspace{0.2cm}
\noindent (3) If follows from Part (1) that, if $\t$ participates in the presence of the CN, then $q_o(\t)\leq q_m(\t).$ Assume now that the inequality $v_o(\t) > v(\t; \pi)$ holds for all $\t$ in a non--empty interval $(\t_1, \t_2)$ and $v_o(\t_1) = v(\t_1; \pi)$ and $v_o(\t_2) = v(\t_2; \pi).$ By the convexity of $v_o$ and $v(\cdot; \pi),$ this would imply the existence of $\t_3\in(\t_1, \t_2)$ such that $v_o'(\t) > v'(\t; \pi)$ holds almost surely in $(\t_1, \t_3).$ However $v_o'(\t)=\psi_1(q_o(\t)),$ $v'(\t; \pi)=\psi_1(q_m(\t))$ and $\psi_1$ is strictly increasing; hence, this would imply that $q_o(\t)>q_m(\t)$ for almost all $\t\in(\t_1, \t_3),$ which is a contradiction.
\hfill $\Box$

\vspace{0.2cm}
We finalize this section with two examples that showcase the results obtained thus far. Example~\ref{MussaRosenEx} showcases that, in the simple case where the outside option is such that the dealer will (only) exclude all high--enough (in absolute value) types, then the results of Theorem~\ref{thm:Main2} follow trivially.

\vspace{0.2cm}

\begin{example}\label{MussaRosenEx}
Let us revisit Example~\ref{MussaRosen} with an extremely steep outside option that will warrant exclusion, namely, for $r_0<r$ let
\be
u_0(\t)=\begin{cases}
				\infty,\text{  if    }\, \t\in [-r,-r_0)\bigcup (r_0, r];\\
					0,\,\,\text{      otherwise}.
				\end{cases}
\ee
Recall that, for a given value $\Gamma$ of the Lagrange multiplier, the corresponding quantity is
\be
q(\t;\Gamma):=2\t+r-2r\Gamma.
\ee
In Example~\ref{MussaRosen} the participation constraint does not bind for high types. In particular, $\gamma\equiv 0$ on $[-r,\underline{\t}_0)$ and to find the left--hand endpoint of the reserved set we set $\Gamma=0$ and solve $2\t+r=0.$ In the current setting, the participation constraint must bind for $\t<-r_0$ and the multiplier will be constant on $(-r_0, \underline{\t}_0(\Gamma)),$ where
\be
\underline{\t}_0(\Gamma):=-\frac{r}{2}\big[1-2\Gamma\big].
\ee
By construction, the choice of $\Gamma$ will bear no weight on the trader types that will be serviced to the left of $\t=-r_0,$ but only on how many additional low types benefit from the presence of the outside option. By integrating $q(\t;\Gamma)$ and noting that the corresponding indirect--utility function $v(\cdot;\Gamma)$ must satisfy $v(\underline{\t}_0(\Gamma);\Gamma)=0,$ we have, for $\t\in [-r_0,\underline{\t}_0(\Gamma)]$
\be
 v(\t;\Gamma)=\t^2+\t r\big[1-2\Gamma\big]+\frac{r^2}{4}\big[1-2\Gamma\big]^2.
\ee
Since the indirect--utility function also satisfies $v(\t;\Gamma)=\t q(\t;\Gamma)-\tau(\t;\Gamma),$ we have that the dealer market on $[-r_0,\underline{\t}_0(\Gamma)]$ is described by the quantity--price pairs $\big(q(\t;\Gamma), \t^2-\frac{r^2}{4}\big[1-2\Gamma\big]^2\big).$ As a consequence, the per--type profit is
\be
\Pi(\t;\Gamma):=-\t^2-\frac{3}{4}r^2\big[1-2\Gamma\big]^2-2\t r\big[1-2\Gamma\big],
\ee
where the third term on the right--hand side is positive and dominates the first two. Finally, we have that each choice of $\Gamma$ will result in the dealer obtaining the aggregate profits from negative types
\be
P(\Gamma):=\frac{1}{2r}\int_{-r_0}^{\underline{\t}_0(\Gamma)}\Pi(\t;\Gamma)d\t.
\ee
The mapping $\Gamma\mapsto P(\Gamma)$ is strictly concave and the first--order conditions yield that it is maximized at $\Gamma=(r-r_0)/(2r).$ As a result $\underline{\t}_0(\Gamma)=-r_0/2$ and
$v(\t;\Gamma)=\t^2+r_0\t+r_0^2/4,$ which correspond to the boundary of the reserved set and the indirect--utility function for negative trader types in the problem without a CN on $[-r_0, r_0].$
\end{example}

\vspace{0.2cm}

\begin{example}\label{RichStructure}
We stay with the basic setup of Examples \ref{ex1} and \ref{ex2}, but now assume that $u_0(\t; \pi)=\Big(\frac{1-  \pi_+}{3}\t^{6/5}-0.001\Big)_+$ for $\t\geq 0$ and $u_0(\t; \pi)\equiv 0$ otherwise.
For any type $\t$ such that $u_0(\t)>0$ it holds that
\be
\big(q_c(\t), \tau_c(\t)\big) =\Big(\frac{2}{5}(1-\pi_+)\t^{1/5},\frac{2}{5}(1-\pi_+)\t^{6/5} + \frac{1}{25}(1-\pi_+)^2\t^{2/5} - \big(\frac13(1-\pi_+)\t^{6/5}-0.001\big)_+\Big).
\ee
We assume $\pi=(0,1/2).$ The first thing to notice is that the dealer's per-type profit for offering $(q_c(\t),\tau_c(\t))$, i.e. $\tau_c(\t)-C(q_c(\t))=\t^{6/5}/30 - \t^{2/5}/100+0.001$, is negative for types $\t\in(0.0035, 0.1667)$. 
On the other hand, the inequality $u_0(\t; 1/2)\geq 0$ only holds for $\t\geq 0.014.$
Combining both arguments we see that $\T_e(\pi)\subset(0.014, 0.1667).$ Next we observe that the inequality
\be
l(\t,1) = K^{-1}\Big(\t-\frac{1 - F(\t)}{f(\t)}\Big)\geq \frac{\sqrt[5]{\t}}{5}
\ee
holds for all $\t\in [0.4761, 1].$ Hence profitable matching may occur on the interval $(0.1667, 0.4761),$ over which $q(\t)=q_c(\t)$ and $\mathbb{C}\big(q(\t)\big)=C\big(q(\t)\big).$ Furthermore, Proposition~\ref{prop:NoJumps2} implies that the corresponding indirect utility function will be differentiable at $\t= 0.4761.$ In order to obtain $v(\t; \pi)$ for $\t\in [0.4761, 1],$ we integrate $l(\cdot, 1)$ and determine the corresponding integration constant $c$ by equating
\be
2\int_0^{0.4761}\left(\frac{3\t^2-6\t+2}{2\t-3}\right)d\t + c = \frac{1}{6}(0.4761)^{6/5} - 0.001.
\ee
We know from the example without a CN that $\gamma(t)=0$ for $\t\in[-1, -0.423).$ On $[-0.423, 0)$ the multiplier must satisfy
\be
K^{-1}\Big(\t-\frac{\gamma(\t) - F(\t)}{f(\t)}\Big)=0,
\ee
which results in $\gamma(\t)=(3\t^2+6\t+2)/4$ on the said interval. What remains to be determined is $\overline{\t}_0$ and $\gamma(\overline{\t}_0).$ To this end, we define the family of functions $v(\cdot;\Gamma)$ such that $v'(\t;\Gamma)=l(\t,\Gamma)$ whenever this quantity is positive and $v(\t;\Gamma)=0$ for $\t\in[0,\t(\Gamma)],$ where $\t(\Gamma)$ is the solution to the equation $l(\t,\Gamma)=0.$ Since $\gamma(0)=0.5,$ we have that $\Gamma>0.5.$\footnote{Pasting when passing from servicing to excluding need not be smooth.}
In fact, $\Gamma=\gamma(\overline{\t}_0)=0.5105,$ $\overline{\t}_0=0.007$ and the intersection of $v(\cdot;\Gamma)$ and $u_0(\cdot; 1/2)$ occurs at $\t= 0.0159.$

Summarizing, the types on $[-1,-0.423)\cup(0.007, 0.0159]\cup(0.1667, 1]$ are fully serviced, those on $[-0.423, 0.007]$ are reserved and the ones that lie on $(0.0159, 0.1667)$ are excluded. The left--hand side of the spread is the same as in the example without a CN, whereas the right--hand side is $t(0_+)=0.0281.$ This is significantly smaller than in Example \ref{ex1}.

\vspace{0.2cm}
Determining $\gamma(\t)$ on $(0, 0.007]$ is relatively simple, as we again must solve $l(\t,\gamma(\t))=0,$ which results in $\gamma(\t)=(-3\t^2+6\t+2)/4$.
Finally, in order to determine $\gamma$ on $\T_e(\pi)$ we must rewrite the virtual surplus using $\mathbb{C}(q(\t))=\tau_c(\t),$ which results in ${\mathbb{C}(q)=(5^5/6)q^6-(1/4)q^2+0.001}.$ The pointwise maximization of the resulting virtual surplus must equal $q_c(\t)=\sqrt[5]{\t}/5.$ After some lengthy arithmetic that we choose to spare the reader from, we obtain
\be
\gamma(\t)=F(\t)-f(\t)\bigg[5^5 q_c(\t)^5-\t\bigg] = F(\t)\quad\text{for }\t\in\T_e(\pi).
\ee
Finally, in the profitable--matching region we solve $l(\t,\gamma(\t))=\sqrt[5]{\t}/5$ so as to find the multiplier, which yields
\be
\gamma(\t)=F(\t)-f(\t)\bigg[\frac{1}{10}\t^{1/5}-\t\bigg]\quad\text{for }\t\in [0.1667,0.4761).
\ee
or
\be
\gamma(\t)=\frac{1}{10}\t^{1/5}\cdot \frac{2\t-3}{4} - \frac{3\t^2-6\t-2}{4} \quad\text{for }\t\in [0.1667,0.4761).
\ee
Observe that, in contrast with Example~\ref{ex2}, here  $\gamma(\t)=1$ for types that are strictly smaller than one. This means that the rightmost types do not profit from introduction of CN via changes in the quantities they are offered, but rather from changes in the corresponding prices. Intuitively speaking this has to do with how steep the outside option is for large types and, as a consequence, whether or not it will be matched over a non-trivial interval.

\vspace{0.2cm}
We present in Figure~\ref{fig:IndUtOpt} the indirect utilities for positive types (the ones for negative ones being the same as in Figure~\ref{fig:IndUt}). The values of $\gamma$ have been plotted in Figure~\ref{fig:OptMult}. In Figure~\ref{fig:WithExZoom} we provide a magnification around small values of $\t$ so as to highlight the switching between reservation, full servicing and exclusion. Observe the jump of the Lagrange multiplier at the boundary between fully--serviced and excluded types (Figure~\ref{fig:OptMultZoom}) and between excluded and matched ones (Figure~\ref{fig:OptMult}).

\begin{figure}[ht!]
\begin{tabular}{cc}
\hspace{-0.5cm}\subfigure[\label{fig:IndUtOpt} Indirect Utilities]
{\includegraphics[width=0.53\textwidth]{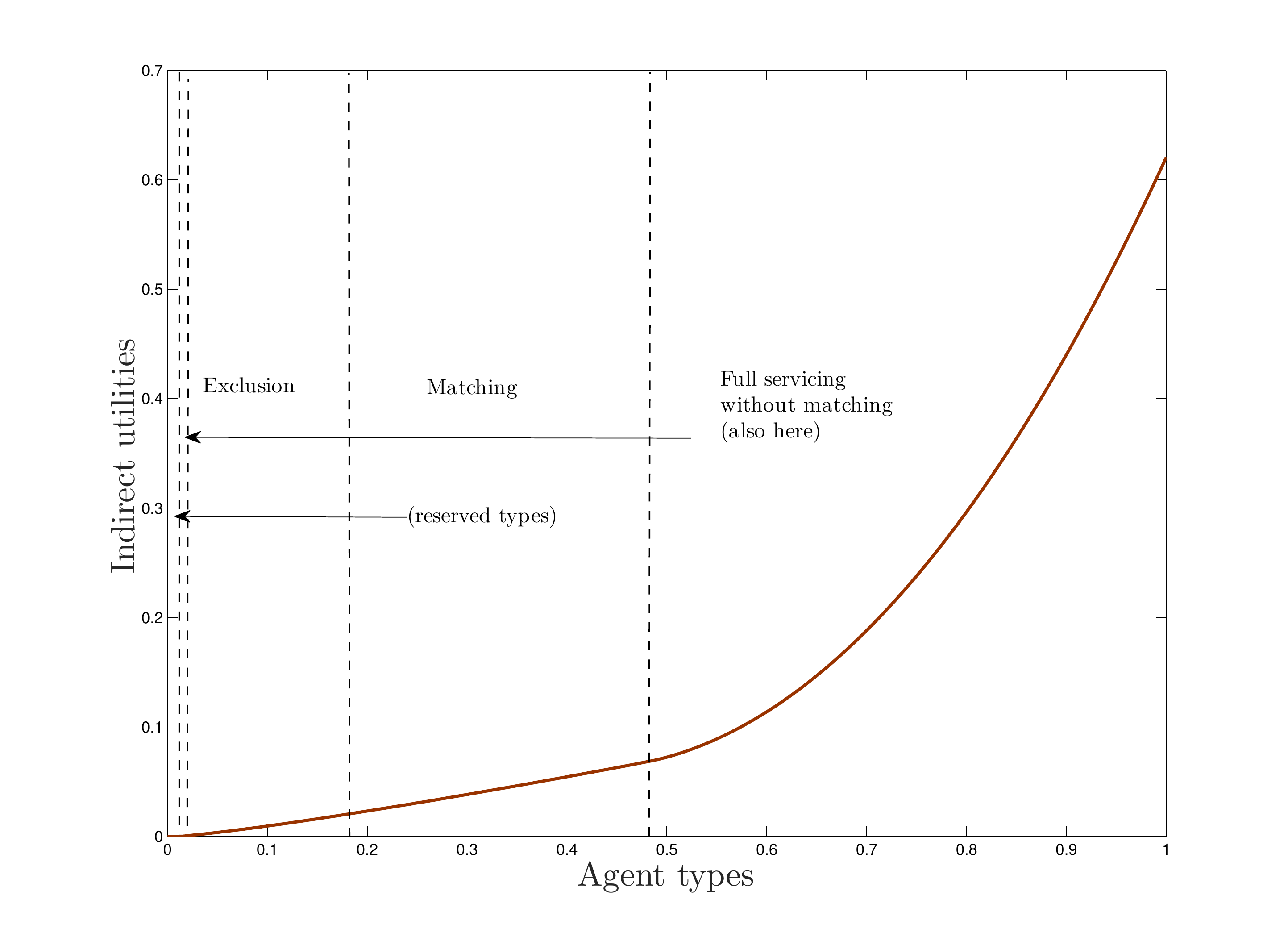}} &
\subfigure[\label{fig:OptMult}Lagrange Multiplier]
{\includegraphics[width=0.52\textwidth]{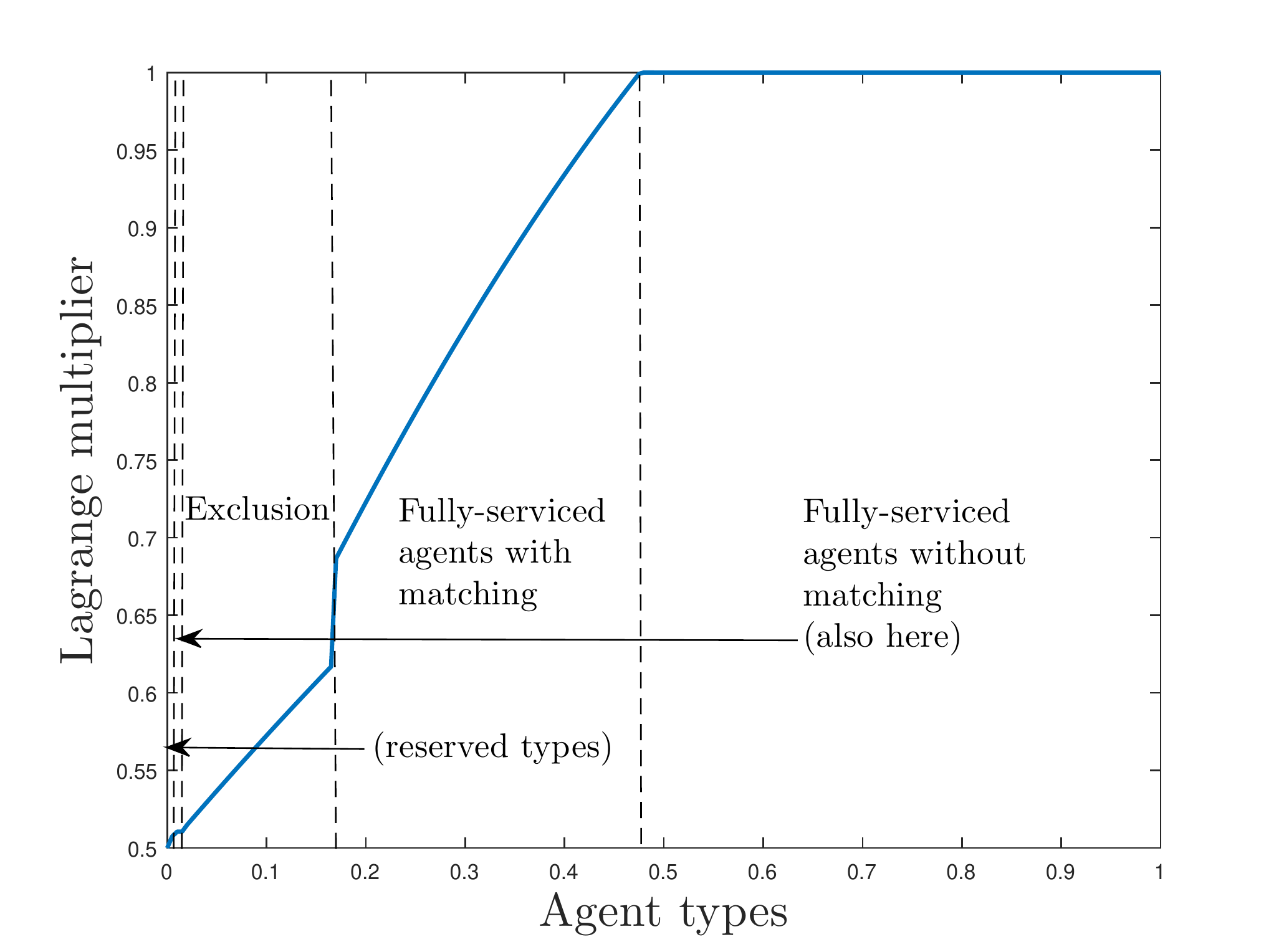}} \\
\end{tabular}\label{fig:WithEx}\caption{An example with exclusion}
\end{figure}

\begin{figure}[ht!]
\begin{tabular}{cc}
\hspace{-0.5cm}\subfigure[\label{fig:IndUtOptZoom} Indirect Utilities]
{\includegraphics[width=0.52\textwidth]{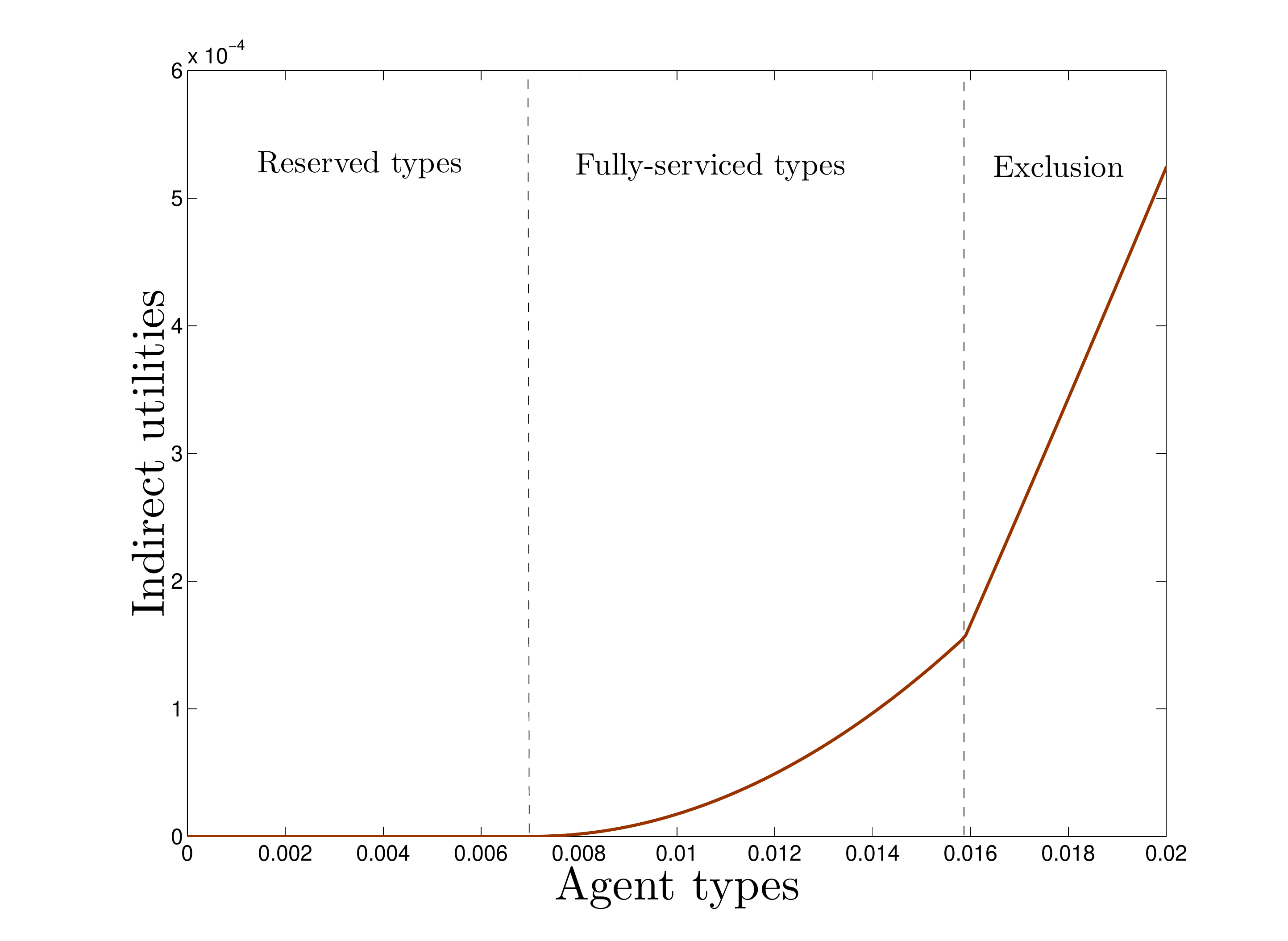}} &
\subfigure[\label{fig:OptMultZoom}Lagrange Multiplier]
{\includegraphics[width=0.52\textwidth]{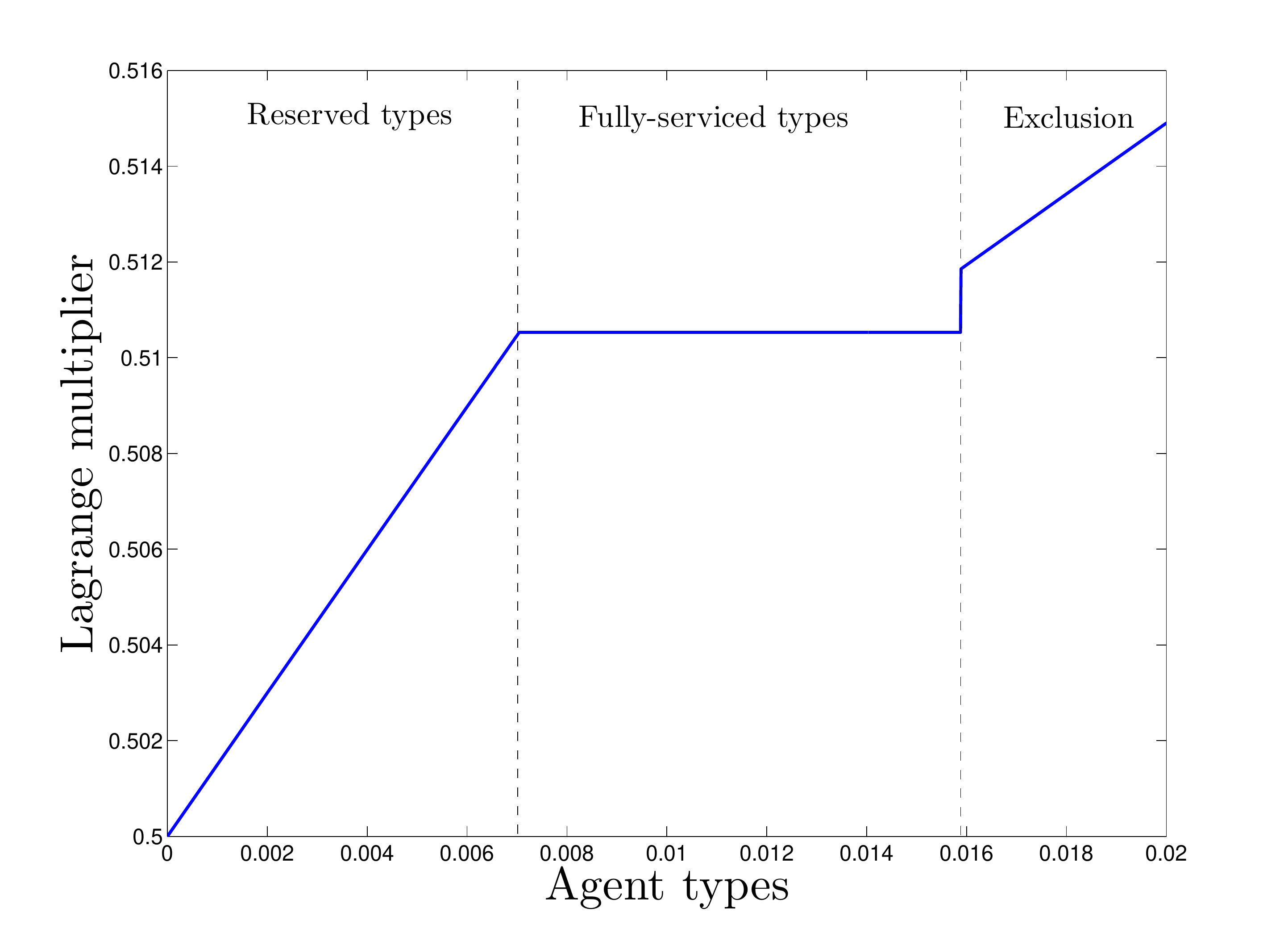}} \\
\end{tabular}\caption{An example with exclusion (magnified)}\label{fig:WithExZoom}
\end{figure}
\end{example}

\noindent We shall revisit this example in the upcoming section, where we look into the existence of equilibrium prices in the CN.

\section{\large{An equilibrium price in the crossing network}}\label{sec:ExistenceEqui}
{In this section we prove the existence of an equilibrium price $\pi^*.$ We first observe that, from Assumption~\ref{ass:monotone}, there is no loss of generality in assuming that $\pi^*$ belongs to some closed and bounded subset of $\re^2,$ which we denote by $\Pi.$ As a consequence we have that $t(0;\cdot):\Pi\to\Pi.$ The restriction of possible equilibrium prices to $\Pi,$ together with Assumptions~\ref{ass:cost of access} and~\ref{ass:monotone}, yields the next result.}

\begin{lemma}\label{lm:BondedParticipation} There exists a non--empty interval $[\epsilon_1, \epsilon_2]\subset\T$ such that
\begin{enumerate}
\item $0\in (\epsilon_1, \epsilon_2);$

\item $u_0(\t; \pi) = 0$ for all $\t\in[\epsilon_1, \epsilon_2]$ and all $\pi\in\Pi.$
\end{enumerate}
\end{lemma}
In the sequel we make use of the results obtained in Section~\ref{sec:ModelCN} to show that the mapping $\pi\mapsto t(0; \pi)$ has the required monotonicity properties so as to use the following result (see, e.g. Aliprantis \& Border~\cite{AB}):

\begin{theorem}\label{thm:Tarski}(Tarski's Fixed Point Theorem) Let $(X, \leq)$ be a non--empty, complete lattice. If $f:X\to X$
is order preserving, then the set of fixed points of $f$ is also a non--empty, complete lattice.
\end{theorem}

\vspace{0.2cm}
\noindent We are now ready to give the proof of our third main result.

\vspace{0.2cm}
\noindent\textbf{Proof of Theorem \ref{thm:Main3}.} Lemmas~\ref{lemma:tradingSB} and ~\ref{lm:BondedParticipation} guarantee that we have a well--defined spread; thus, we may decompose the analysis of the mapping $\pi\mapsto t(0; \pi)$ into that of the mappings $\pi_-\mapsto t(0_-; \pi_-)$ and $\pi_+\mapsto t(0_+; \pi_+).$ In other words, for a given price $\pi,$ the dealer's optimal response to $u_0(\cdot;\pi)$ is, modulo a normalization of $\gamma,$ equivalent to the combination of his actions towards negative and positive types separately. We shall concentrate on the existence of a fixed point of the mapping $\pi_+\mapsto t(0_+; \pi_+).$

\vspace{0.2cm}
From Assumption~\ref{ass:monotone} we have that if $\pi_{1+}<\pi_{2+}$, then $u_0(\t;\pi_{1+})>u_0(\t;\pi_{2+})$ for all $\t>0.$ If for $i=1,2$ it holds that $u_0(\t;\pi_{i+})<v_o(\t)$ for all $\t>0$, then $v(\t;\pi_{1+})=v(\t;\pi_{2+})$ on the same domain and $t(0_+; \pi_{1+})=t(0_+; \pi_{2+}).$ Next assume that $u_0(\t;\pi_{i+})\geq v_o(\t)$ on a subset $\T_i$ of $(0, \overline{\t}],$ for $i=1,2.$  Given that $u_0(\t;\pi_{1+})>u_0(\t;\pi_{2+})$ for all $\t>0,$ then $\overline{\t}(\pi_1)<\overline{\t}(\pi_2)$ and the first point $\widetilde{\t}_1$ such that $v(\t;\pi_{1+})=u_0(\t;\pi_{1+})$ holds satisfies $\widetilde{\t}_1<\widetilde{\t}_2,$ where the latter is the analogous to $\widetilde{\t}_1$ in the presence of $u_0(\t;\pi_{2+}).$ The existence of $\widetilde{\t}_1$ and $\widetilde{\t}_2$ is guaranteed by the fact that in both cases the indirect--utility functions intersect the corresponding outside options. Arguing as in the proof of Theorem~\ref{thm:Main2}, Part (2), this also implies that $\overline{\t}_0(\pi_{1})<\overline{\t}_0(\pi_{2});$ hence $t(0_+; \pi_{1+})<t(0_+; \pi_{2+}).$ In other words, the mapping $\pi_{+}\mapsto t(0_+; \pi_{+})$ is order--preserving and, using Tarski's Fixed Point Theorem, we may conclude it has a fixed point. \hfill $\Box$

\vspace{0.2cm}

\begin{remark}
The requirement of uniformly distributed types can be relaxed to the extent that if $f$ and $K$ are such that Conditions~\eqref{eq:condMonot} are satisfied, then the required monotonicity properties still apply. Unfortunately, these conditions cannot be verified ex--ante, since they include the end points of the set of reserved traders.
\end{remark}

\vspace{0.2cm}

\begin{example}
Let us go back to our example with exclusion, but introduce the feedback loop between the DM and the CN through the iteration $\pi_{i+1}= t(0;\pi_i).$ We initialize the recursion by setting $\pi_0=(0, 1/2)$ and $\kappa=0.001$, which are the parameters in the aforementioned example.

\begin{figure}[ht!]
	\centering
		\includegraphics[width=0.6\textwidth]{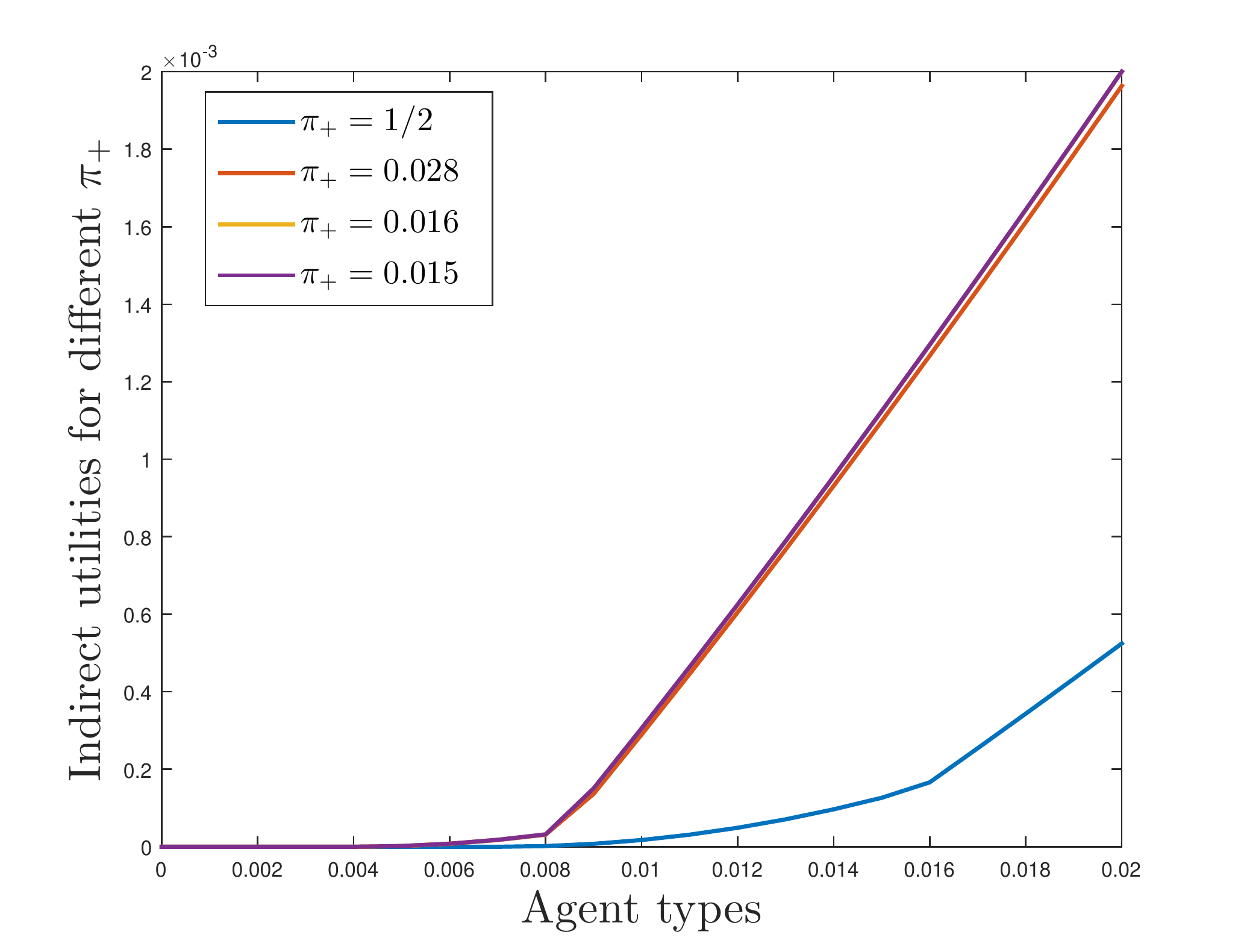}
	\caption{
The indirect--utility functions corresponding to the iteration $\pi_{i+1}=t(0; \pi_{i})$.\label{fig:EquilPriceZoomIn}}
\end{figure}

\vspace{0.2cm}
We observe a very swift convergence. Indeed, it takes only four iterations to reach $\|v(\cdot;\pi_i)-v(\cdot; \pi_{i+1})\|_{\infty}\leq 10^{-5}$ and the indirect--utility functions in the third and fourth iteration are almost indistinguishable. The equilibrium price is $\pi^*=(0, 0.015)$. We present in Figure~\ref{fig:EquilPriceZoomIn} the plots of the first four iterates. It is evident that each iteration results in a smaller set of reserved traders and in a higher indirect utility for all types. The spreads, the right endpoints of the reserved regions, the Lagrange multipliers at the right endpoint of the reserved regions and the exclusion regions are provided in Table~\ref{table1}. It is interesting to observe that, as the spread decreases to its equilibrium level, the number of trader types that are reserved decreases and the sets of excluded types grow (in terms of inclusions). This last fact obeys the fact that, when the traders have a more attractive outside option, it is harder for the dealer to match it profitably.

\begin{table*}[ht!]
\small
\centering
\caption{The numbers of the feedback loop}\label{table1} \vspace{0.5cm}
\begin{tabular}{@{}ccc ccc cccc cccc @{}}\toprule
	 $\pi_+$  &   		$\T_o$			 & $\Gamma$&	     $\T_e(\pi_+)$	 \\ \midrule

	 $ 1/2$   &	  [-0.423,0.0070]  &	 0.5105  &	[0.0159, 0.1667]\\
	$0.0281$  &	  [-0.423,0.0040]  &	 0.5061  &	[0.0083, 0.4872]\\
	$0.0161$  &	  [-0.423,0.0040]  &	 0.5060  &	[0.0082, 0.4954]\\
	$0.0158$  &	  [-0.423,0.0040]  &	 0.5060  &	[0.0082, 0.4955]\\
\bottomrule \\
\end{tabular}
 \end{table*}
\end{example}

\section{\large{Portfolio liquidation and dark--pool trading}}\label{sec:DPtrading}

{In this section we present an application of our methodology to portfolio liquidation. We assume that the market participants' aim is to liquidate their current holdings on some traded asset. The sizes of the traders' portfolios are heterogeneous and saying that a trader's type is $\t$ means that he holds $\t$ shares of the asset prior to trading.} We set $\T=[-1, 1]$ and $f\equiv 1/2.$ If a trader of type $\t$ trades $q$ shares for $\tau$ dollars, his utility is
\be
\hat{u}(\t, q)- \tau:=-\alpha(\t-q)^2-\tau,
\ee
where $0<\alpha$ denotes the traders' (homogeneous) sensitivity towards inventory holdings. Notice that $-\alpha\t^2$ is the type--dependent reservation utility of a trader of type $\t.$ If we ``normalize" the said utility to zero, we may write
\be
u(\t ,q) - \tau = 2\alpha\t q -\alpha q^2 -\tau.
\ee
In this example the crossing network takes the form of a \textit{dark pool} (DP). Choosing to trade in the latter entails two kinds of costs for the traders: On the one hand, there is a direct, fixed cost $\kappa>0$ of engaging in dark--pool trading. On the other hand, execution in the DP is not guaranteed. We denote by $p\in [0, 1]$ the probability that an order is executed where we assume for simplicity that the probability of order execution is independent of the order size. Pricing in the DP is linear. Namely, for a given execution price $\pi$, the utility that a trader of type $\t$ extracts from submitting an order of $q$ shares to be traded in the DP is
\be
p\big[(2\t\alpha - \pi)q - \alpha q^2\big] - \kappa,
\ee
where again we have normalized reservation utilities to zero. The problem of optimal submission to the DP for a $\t$--type trader is
\be
\max_q \Big\{p\big[(2\t\alpha - \pi)q - \alpha q^2\big]\Big\},
\ee
which yields the optimal submission level
\be
q_d(\t):= \t-\frac{\pi}{2\alpha}.
\ee
We obtain that opting for the DP results in a trader of type $\t$ enjoying the expected utility
\be
u_0(\t;\pi) = \alpha p\left(\t-\frac{\pi}{2\alpha}\right)^2 - \kappa.
\ee
We assume that $p\pi^2<4\alpha\kappa$ so as to keep the DP unattractive for small types.

\vspace{0.2cm}
We assume that the dealer's costs/profits of unwinding a portfolio of size $q$ are $C(q)=\epsilon\,q+\beta q^2$ where $\beta>0$ and $\epsilon$ is non--negative. Observe that, since
$u_0(\cdot;\pi)$ does not satisfy Assumption~\ref{ass:cost of access}, some restrictions must be imposed on the problem's parameters so as to still have Lemma~\ref{lm:BondedParticipation}.
Namely, it must hold that
\begin{equation}\label{eq:RestParamDP}
\pi<2\sqrt{\frac{\alpha\kappa}{p}}.
\end{equation}
Condition~\eqref{eq:RestParamDP} imposes a hard upper bound on possible equilibrium DP prices. It should be noted that Assumption~\ref{ass:monotone} is not satisfied by $u_0(\cdot;\pi),$ which, together with the way in which we shall define the pricing feedback loop from the DM to the DP, implies that our equilibrium result does not apply ``as is'' to the current setting.

\subsection{\large{The dealer market without a dark pool}}

In the absence of a dark pool, the dealer's optimal choices of quantities are, for negative types
\be
l(\t, 0)=\frac{\alpha}{\alpha+\beta}\big(2\t+1\big)-\frac{\epsilon}{2(\alpha+\beta)}
\ee
and for positive types
\be
l(\t, 1)=\frac{\alpha}{\alpha+\beta}\big(2\t-1\big)-\frac{\epsilon}{2(\alpha+\beta)},
\ee
where the boundary of $\T_0$ is given by
\be
\underline{\t}_0=\frac{1}{2}\Big(\frac{\epsilon}{2\alpha}-1\Big)\quad\text{and}\quad \overline{\t}_0=\frac{1}{2}\Big(\frac{\epsilon}{2\alpha}+1\Big).
\ee
In order to guarantee that $\T_0\subset[-1,1]$ the condition $\epsilon<2\alpha$ must be imposed on the corresponding parameters. From the relation $v'(\t)=\Psi_1\big(q(\t)\big)$ we have that the indirect--utility function is
\be
v(\t)=\begin{cases}
					\frac{2\alpha^2}{\alpha+\beta}\t^2 +\frac{\alpha}{\alpha+\beta}\big(2\alpha-\epsilon\big)\t+c_1, & \t\leq\underline{\t}_0;\\
					\frac{2\alpha^2}{\alpha+\beta}\t^2 -\frac{\alpha}{\alpha+\beta}\big(2\alpha+\epsilon\big)\t+c_2, & \t\geq\overline{\t}_0,
					\end{cases}
\ee
where
\be
c_1=\frac{2\alpha^2}{4(\alpha+\beta)}\Big(\frac{\epsilon}{2\alpha}+1\Big)^2 +\frac{\alpha(2\alpha+\epsilon)}{2(\alpha+\beta)}\Big(\frac{\epsilon}{2\alpha}+1\Big)\text{  and  } c_2=\frac{2\alpha^2}{4(\alpha+\beta)}\Big(\frac{\epsilon}{2\alpha}-1\Big)^2 -\frac{\alpha(2\alpha+\epsilon)}{2(\alpha+\beta)}\Big(\frac{\epsilon}{2\alpha}-1\Big).
\ee
When it comes to the spread, observe that $q'\equiv \frac{2\alpha}{\alpha+\beta},$ $\psi_1\equiv 2\alpha$ and $\psi_2\equiv0,$ which yields
\be
[t(0_-), t(0_+)]=\frac{4\alpha^2}{\alpha+\beta}[\underline{\t}_0,\overline{\t}_0].
\ee
Below we analyze how the spread changes with the introduction of the DP.

\subsection{\large{The impact of a dark pool}}

We first take an exogenous execution price $\pi$ and determine, for each $\t\in\T,$ what is the quantity--price pair $\big(q_c(\t;\pi), \tau_c(\t;\pi)\big)$ that the dealer must offer so as to match a DP with execution price $\pi.$ Using the relation $q_c(\t;\pi)=u_0'(\t;\pi)$ we obtain
\begin{align}\begin{split}
q_c(\t;\pi)=& \ 2\alpha p\left(\t-\frac{\pi}{2\alpha}\right) \text{ and}\\
\tau_c(\t;\pi)=& \ \kappa+4\alpha^2 p(\t -\alpha p)\left(\t-\frac{\pi}{2\alpha}\right)-\alpha p\left(\t-\frac{\pi}{2\alpha}\right)^2.
\end{split}\end{align}

From the Envelope Theorem and the structure of $u(\t,q)$ we have that the traders' indirect utility function satisfies
\begin{equation}\label{eq:QualEnvel}
\frac{v'(\t)}{2\alpha}=l\big(\t,\gamma(\t)\big).
\end{equation}
In order to determine the spread in the presence of the DP we must determine $\underline{\t}_{0, m}$ and $\overline{\t}_{0, m}$ together with $\gamma\big(\underline{\t}_{0, m}\big)$ and $\gamma\big(\overline{\t}_{0, m}\big).$ For an arbitrary $\Gamma\in [0, 1]$ we have
\be
l(\t,\Gamma)=\frac{\alpha}{\alpha+\beta}\big[2\t+1-2\Gamma\big]-\frac{\epsilon}{2(\alpha+\beta)}.
\ee
Indexed by $\Gamma,$ the candidates for $\underline{\t}_{0, m}$ are then given by
\be
\underline{\t}_{0, m}(\Gamma)=\frac{1}{2}\Big(\frac{\epsilon}{2\alpha}+2\Gamma-1\Big).
\ee
Since it must hold that $\underline{\t}_{0, m}(\Gamma)\leq 0$, then $\Gamma\leq 0.5(1-\epsilon/2\alpha).$ Integrating Expression~\eqref{eq:QualEnvel} we have that, on the interval $[\widetilde{\t}_{m}(\Gamma), \underline{\t}_{0, m}(\Gamma)],$ the traders' indirect utility is given by
\begin{equation}\label{eq:IndUtGamma}
v(\t;\Gamma)=\frac{2\alpha^2}{\alpha+\beta}\t^2+2\alpha\Big[\frac{\alpha}{\alpha+\beta}(1-2\Gamma)-\frac{\epsilon}{2(\alpha+\beta)}\Big]\t+c_{1,m},
\end{equation}
where $\widetilde{\t}_{m}(\Gamma)$ is the first intersection to the left of $\underline{\t}_{0, m}(\Gamma)$ of $v(\cdot; \Gamma)$ and $u_0(\cdot;\pi)$ and $c_{1,m}$ is determined by the equation
\be
v\big(\underline{\t}_{0, m}(\Gamma);\Gamma\big)=0.
\ee
Unless the inequality $\Gamma\leq 0.5(1-\epsilon/2\alpha)$ is tight, in which case the types below $\widetilde{\t}_{m}(\Gamma)$ are excluded, Proposition~\ref{prop:NoJumps2} implies that $\Gamma$ must be chosen so as to satisfy the smooth--pasting condition
$u_0'\big(\widetilde{\t}_{m}(\Gamma);\pi\big)=v'\big(\widetilde{\t}_{m}(\Gamma);\pi\big),$ which is equivalent to
\be
\widetilde{\t}_{m}(\Gamma)=\Big[\frac{2\alpha}{\alpha+\beta}-p\Big]^{-1}   \Big[\frac{\epsilon}{2(\alpha+\beta)}-\frac{\alpha}{\alpha+\beta}(1-2\Gamma)-\frac{p\pi}{2\alpha}\Big].
\ee
Observe that, besides the requirement $\Gamma\geq 0.5(1-\epsilon/2\alpha),$ the strategy to determine $\overline{\t}_{0, m}$ is exactly the same as for $\underline{\t}_{0, m}.$ Summarizing, from Eq.~\eqref{eq:IndUtGamma} we observe that, if $\Gamma_-$ and $\Gamma_+$ correspond to the optimal choices for the negative and positive endpoints of $\T_{0}(\pi),$ then
\be
q'\big(\underline{\t}_{0, m}(\Gamma_-)\big)=\frac{1}{2\alpha}v''\big(\underline{\t}_{0, m}(\Gamma_-);\Gamma_-\big)=
\frac{1}{2\alpha}v''\big(\overline{\t}_{0, m}(\Gamma_+);\Gamma_+\big)=q'\big(\overline{\t}_{0, m}(\Gamma_+)\big)=\frac{2\alpha}{\alpha+\beta}.
\ee
The spread is then
\be
[t_m(0_-), t_m(0_+)]=\frac{4\alpha^2}{\alpha+\beta}[\underline{\t}_{0, m}(\Gamma_-),\overline{\t}_{0, m}(\Gamma_+)]\subset \frac{4\alpha^2}{\alpha+\beta}[\underline{\t}_0,\overline{\t}_0],
\ee
i.e. the presence of a dark pool strictly narrows the spread in the dealer's market.

\subsection{\large{An equilibrium price}}
A standard (but not unique) way in which dark--pool prices are generated is by computing the average of some publicly available best--bid and best--ask prices. In the case of the US, this is usually the mid--quote of the National Best Bid and Offer (NBBO). Borrowing from this idea we define the price--iteration in the DP as follows:
\be
\pi_{i+1}=\frac{1}{2}\big(t_i(0_+)-t_i(0_-)\big),\quad i\in\n,
\ee
where $\{t_i(0_-), t_i(0_+)\}$ are the best bid and ask prices in the DM in the presence of a DP with execution price $\pi_i.$ We know from the previous section that the sequence $\{\pi_i, i\in\n\}\subset ((4\alpha^2)/(\alpha+\beta))[\underline{\t}_0,\overline{\t}_0];$ hence, by the Bolzano--Weierstrass Theorem it has at least one convergent subsequence. The limit of each of the said subsequences will be an equilibrium price. The (possible) non--uniqueness of these prices is due to the fact that by virtue of its definition, the sequence of dark--pool prices need not be monotonic. The problem of non-uniqueness of equilibria in models of competing DMs and CNs has been observed before. We refer to \cite{DDH} for a detailed discussion.

\section{\large{Conclusions}}\label{sec:Conclusions}

We have presented an adverse--selection model to study the structure of the limit--order book of a dealer who provides liquidity to traders of unknown preferences. Furthermore, we have established a link between the traders' indirect--utility function and the bid--ask spread in the DM. Making use of the aforementioned link, we have studied how the presence of a type--dependent outside option impacts the spread of the DM, as well as the set of trader types who participate in the DM and their welfare. In particular, we have shown, in a portfolio--liquidation setting, that the presence of a dark pool results in a shrinkage of the spread in the DM. Finally, we have established that, under certain conditions, the feedback loop introduced by the impact that the spread has on the structure of the outside option leads to an equilibrium price.

\bibliographystyle{elsarticle-harv}
\bibliography{CNBHM-Arxiv}
\end{document}